\documentclass[aps,twocolumn,english,superscriptaddress,nofootinbib]{revtex4-1}
\usepackage[utf8]{inputenc}

\usepackage{amsmath}
\usepackage{appendix}
\usepackage[english]{babel}
\usepackage{geometry}
\usepackage{graphicx}
\usepackage{hyperref,cleveref}
\usepackage{microtype}
\usepackage{multirow}
\usepackage[binary-units=true,exponent-product=\cdot]{siunitx}
\usepackage{tikz} 

\usepackage{texmf/abbreviations}
\usepackage{texmf/formulautils}

\hypersetup{
	colorlinks = true,
	linkcolor  = cyan,
	urlcolor   = blue,
	breaklinks = true
}

\crefname{equation}{Eq.}{Eqs.}
\Crefname{equation}{Eq.}{Eqs.}

\crefname{section}{Sec.}{Secs.}
\Crefname{section}{Sec.}{Secs.}

\crefname{figure}{Fig.}{Figs.}
\Crefname{figure}{Fig.}{Figs.}

\NewDocumentCommand{\hsprod}{O{}mm}{#1\langle#2,#3#1\rangle_\mathrm{HS}}

\DeclareMathOperator*{\avg}{avg}
\DeclareMathOperator*{\argmax}{argmax}
\DeclareMathOperator*{\Span}{span}

\NewDocumentCommand{\quant}{m}{
	\IfEqCase{#1}{%
		{fidelity}{\mathcal{F}}%
		{non_leakage}{\mathcal{F}'}%
	}%
	[\PackageError{}{Undefined quantity:}{}]%
}

\begin{document}

\title{Efficient cavity control with SNAP gates}

\author{Thomas Fösel}
\affiliation{Max Planck Institute for the Science of Light, Staudtstr. 2, 91058 Erlangen, Germany}
\affiliation{Department of Applied Physics and Physics, Yale University, New Haven, CT, USA}
\affiliation{Yale Quantum Institute, Yale University, New Haven, CT, USA}

\author{Stefan Krastanov}
\affiliation{Department of Applied Physics and Physics, Yale University, New Haven, CT, USA}
\affiliation{Yale Quantum Institute, Yale University, New Haven, CT, USA}
\affiliation{Department of Electrical Engineering and Computer Science, Massachusetts Institute of Technology, Cambridge, MA 02139, USA}
\affiliation{John A.\ Paulson School of Engineering and Applied Sciences, Harvard University, Cambridge, MA 02138, USA}

\author{Florian Marquardt}
\affiliation{Max Planck Institute for the Science of Light, Staudtstr. 2, 91058 Erlangen, Germany}
\affiliation{Physics Department, University of Erlangen-Nuremberg, Staudstr. 5, 91058 Erlangen, Germany}

\author{Liang Jiang}
\affiliation{Department of Applied Physics and Physics, Yale University, New Haven, CT, USA}
\affiliation{Yale Quantum Institute, Yale University, New Haven, CT, USA}
\affiliation{Pritzker School of Molecular Engineering, The University of Chicago, Chicago, Illinois 60637, USA}

\begin{abstract}
	Microwave cavities coupled to superconducting qubits have been demonstrated to be a promising platform for quantum information processing. A major challenge in this setup is to realize universal control over the cavity. A promising approach are selective number-dependent arbitrary phase (SNAP) gates combined with cavity displacements. It has been proven that this is a universal gate set, but a central question remained open so far: how can a given target operation be realized efficiently with a sequence of these operations. In this work, we present a practical scheme to address this problem. It involves a hierarchical strategy to insert new gates into a sequence, followed by a co-optimization of the control parameters, which generates short high-fidelity sequences. For a broad range of experimentally relevant applications, we find that they can be implemented with $3$ to $4$ SNAP gates, compared to up to $50$ with previously known techniques.
\end{abstract}

\maketitle

\section{Introduction}

In the past decade, remarkable progress has happened in the field of quantum information processing \cite{nielsen2011book,ladd2010review_qc}. One of the leading platforms is circuit QED with microwave cavities coupled to superconducting qubits. While the (three-dimensional) cavities offer relatively long coherence times of around $1\dots\SI{10}{\milli\second}$, they allow only for very restricted direct control via cavity displacements (driven with a microwave pulse). In contrast, the non-linearity of superconducting qubits like transmons \cite{schoelkopf2008review_superconducting_qubits,devoret2013review_superconducting_qubits,wendin2017superconducting_qubits} allows for universal control over them, but their coherence time is much shorter (up to $\SI{100}{\micro\second}$). To combine the respective strengths of these systems, \abbr{ie} the long lifetime and the good controllability, the quantum information can be stored in (a single mode of) the cavity, and the transmon is used as an ancilla to manipulate the cavity state. Impressive recent demonstrations using this platform are the realization of quantum error correction beyond the breakeven point \cite{ofek2016cat_code_experiment}, entanglement between remote systems \cite{axline2018on_demand} and deterministic teleportation of quantum gates \cite{chou2018deterministic_teleportation}.

Currently, there are two competing approaches towards achieving universal control over the cavity with the help of a transmon. The first one makes use of the full freedom of the control space in order to find suitable control pulses \cite{heeres2017grape}, which is typically done using numerical optimal-control techniques like gradient ascent pulse engineering (GRAPE \cite{khaneja2005grape}). The second approach is to build combinations from a restricted set of well-controllable gates. A promising candidate for such a gate set is the selective number-dependent arbitrary phase (SNAP) gate \cite{heeres2015snap_experiment} combined with cavity displacements, for which universality has been proven \cite{krastanov2015snap_theory}. The main advantage of this second approach is its modularity, which results in better interpretability and simplifies the optimization of its constituents. In particular, the SNAP gate can be made robust against transmon decay via error transparency \cite{reinhold2019snap_error_transparency,ma2019path_independent_gates}.

One central challenge in this gate-based approach is to find suitable combinations for the (continuous) control parameters of these gates, which represents a complex task due to the huge size of the search space. Furthermore, the generated gate sequences have to be efficient, \abbr{ie} the number of gates needs to be small, as the following consideration shows.

The previously best known technique from \cite{krastanov2015snap_theory} generates sequences with (at least) $N^2$ SNAP gates interleaved with $N^2+1$ displacements to implement an operation on $N$ different Fock states. \Abbr{eg}, for $N=10$, this means that $100$ SNAP gates would be required. With $3\hdots\SI{5}{\micro\second}$ per SNAP gate and less than $\SI{0.1}{\micro\second}$ per displacement \cite{axline2018on_demand}, the total execution time would be $300\hdots\SI{500}{\micro\second}$, resulting in significant corruption of the quantum information given the lifetime of the cavity ($1\dots\SI{10}{\milli\second}$).

A simple parameter-counting argument suggests that the actually required number of gates should be significantly lower. We compare the number of conditions to match an operation on $N$ Fock states, which is $\dim(\mathset{SU}(N))=N^2-1=\mathcal{O}(N^2)$, to the effective number of free parameters for each pair of displacement and SNAP gate, which is $\mathcal{O}(N)$. From the quotient of these values, we see that it should be possible to realize such an operation already within $\mathcal{O}(N)$ gates. This would be a quadratic speedup over the technique from \cite{krastanov2015snap_theory}. For details, see \cref{sec:appendix:gate_seq:length_estimate}.

In this work, we present a practical scheme to determine the control parameters for sequences of SNAP gates and displacements to match a given target operation. The results indicate that our scheme indeed follows this linear scaling. Further, we apply our technique to optimize several experimentally relevant applications. We find that, for example, correcting errors in a cavity code can be performed with 4 SNAP gates (compared to up to 50 with the old technique). Therefore, our work promises to increase the fidelity for operations in circuit QED and thereby paves the way for novel experiments.

\begin{figure}[t!]
	\centering
	\includegraphics{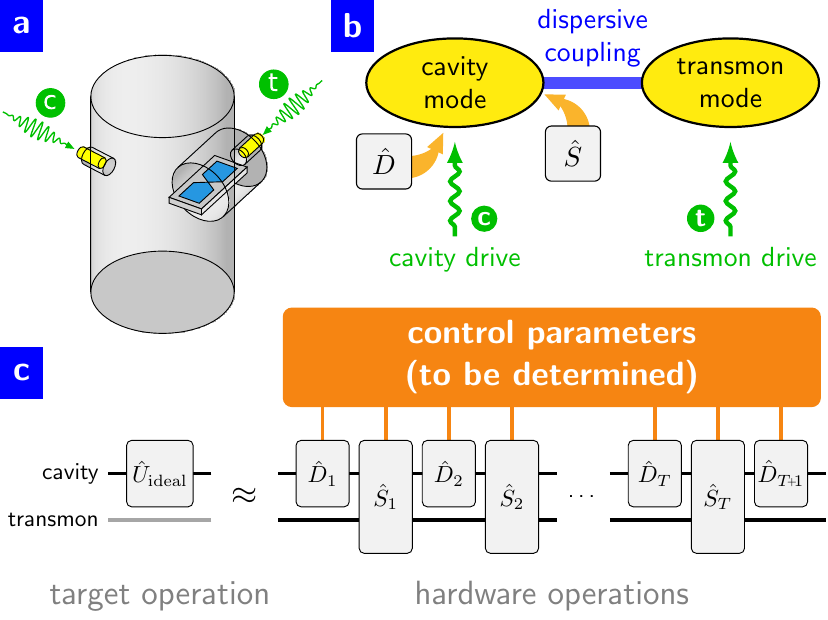}
	\caption{Setup and goal.
		a) Circuit QED platform: a microwave cavity (gray) coupled to a transmon qubit (blue).
		b) Modes and drives. The cavity and the transmon each contribute one mode to the system. The cavity mode is used to store the quantum information. The cavity drive allows to perform displacements $\hat{D}$ on the cavity mode. Via the dispersive coupling, also the transmon drive indirectly affects the cavity mode; in this application, this is used to implement SNAP gates \cite{heeres2015snap_experiment,krastanov2015snap_theory} with the action $\hat{S}(\vec{\theta})=\sum_{n=0}^{\infty}\mconst{e}^{\mconst{i}\theta^{(n)}}\Ketbra{n}$ on the cavity.
		c) Universal control schemes. An arbitrary target operation $\hat{U}_\mathrm{ideal}$ can be approximately decomposed into an alternate series of SNAP gates $\hat{S}$ and displacements $\hat{D}$, as guaranteed by an existence theorem \cite{krastanov2015snap_theory}. We present a practical construction scheme for such gate sequences, including their relevant control parameters. We try to achieve high fidelity while minimizing the length of the gate sequence (quantified by $T$, the number of SNAP gates).
	}
	\label{fig:setup}
\end{figure}

\section{Problem specification}
\label{sec:problem_spec}

Our goal is to (approximately) decompose a given target operation into a series of hardware operations. In our circuit QED setup, these hardware operations are SNAP gates and displacements. As a practical way to find gate sequences with high fidelity under realistic experimental conditions, we aim to achieve a good trade-off between the following three objectives: \textit{(i)} a large overlap between actual and desired output states disregarding noise and gate imperfections, \textit{(ii)} a small number of gates, \abbr{ie} an efficient sequence, and \textit{(iii)} low photon number at intermediate times. In this section, we will motivate this approach and make these criteria precise.

We start by introducing the two native hardware operations: SNAP gates $\hat{S}(\vec{\theta})$ and displacements $\hat{D}(\alpha)$. In this work, we will not address their hardware implementation (see \cite{heeres2015snap_experiment,krastanov2015snap_theory,reinhold2019snap_error_transparency} for further reading), but work directly with their resulting action on the cavity:
\begin{align}
	\hat{S}(\vec{\theta}) & = \sum_{n=0}^{\infty} \mconst{e}^{\mconst{i}\theta^{(n)}} \Ketbra{n} &
	\hat{D}(\alpha) & = \exp(\alpha\hat{a}^\dagger+\conj{\alpha}\hat{a}).
	\label{eq:def_snap_and_displacment}
\end{align}
These operators are idealized in the sense that noise and gate imperfections are not taken into account directly. However, we indirectly tackle these effects by searching for gate sequences with small number of gates and low photon number.

The interesting sequences are those which interleave the two gate types, as equal pairs could always be contracted to a single gate. This leads to the ansatz
\begin{equation}
	\hat{U} = \hat{D}(\alpha_{T+1}) \cdot \hat{S}(\vec{\theta}_T) \cdot \hat{D}(\alpha_T) \cdot \hdots \cdot \hat{S}(\vec{\theta}_2) \cdot \hat{D}(\alpha_2) \cdot \hat{S}(\vec{\theta}_1) \cdot \hat{D}(\alpha_1).
	\label{eq:def_gate_sequence}
\end{equation}
Here, the integer $T$ counts the number of SNAP gates and thereby characterizes the length of the sequence. While \cref{eq:def_gate_sequence} determines the sequence structure, we still need to find explicit choices for the control parameters, $\alpha_t$ and $\vec{\theta}_t$, of the gates in the sequence.

The action of the target operation does not need to be defined on the entire Hilbert space $\mathset{H}$, but can be restricted to a logical subspace of dimension $L$. In the simplest case, we might only care about a single state $\ket{\psi_\mathrm{in}}$ which should be transformed into another state $\ket{\psi_\mathrm{out}}$. In this case, we have $L=1$. As we can also see from this example, the input and output space do not have to coincide; here, it changes from $\Span(\ket{\psi_\mathrm{in}})$ to $\Span(\ket{\psi_\mathrm{out}})$.

In general, we have a (pre-determined) target operation defining an isometric mapping from an input space $\mathcal{X}\subseteq\mathset{H}$ to an output space $\mathcal{Y}\subseteq\mathset{H}$. We fix an orthonormal basis $\{\ket{x_1},\hdots,\ket{x_L}\}\subset\mathcal{X}$ of the input space, and its image $\{\ket{y_1},\hdots,\ket{y_L}\}\subset\mathcal{Y}$ \abbr{wrt} the target operation, which is a basis of the output space. With this, we can define the operator
\begin{equation}
	\hat{V} := \sum_{l=1}^{L} \ketbra{y_l}{x_l}
	\label{eq:definition_target_operation}
\end{equation}
to uniquely characterize the target operation ($\hat{V}$ does not depend on the choice of the basis).

For $\hat{U}$, we are restricted to unitary transformations, but $\hat{V}$ itself is not unitary unless the target operation is defined on the entire Hilbert space. However, $\hat{V}$ can always be extended to unitary transformations which satisfy
\begin{equation}
	\hat{U}_\mathrm{ideal} \ket{x} = \hat{V} \ket{x} \in \mathcal{Y}
\end{equation}
for all $\ket{x}\in\mathcal{X}$. The transformation $\mconst{e}^{\mconst{i}\varphi}\hat{U}_\mathrm{ideal}$ would perform the target operation perfectly, where $\varphi$ represents a possible global phase.

In practice, it might not be possible to implement $\mconst{e}^{\mconst{i}\varphi}\hat{U}_\mathrm{ideal}$ precisely. Even assuming ideal gates, as in \cref{eq:def_snap_and_displacment}, most target operations cannot be matched exactly with a (finite) sequence of hardware operations. This error is addressed by objective \textit{(i)}. In a real experiment, noise and gate imperfections will further decrease the fidelity. Because these effects accumulate with each new gate, their error decreases for shorter sequences, which leads us to demand objective \textit{(ii)}. In addition, also the mean photon number matters, as this determines the rate of photon losses, the dominant noise process for a cavity. The photon number can change under displacements. So, even though the initial and final states are fixed, higher Fock states can get populated at intermediate points of the sequence. This cannot completely be avoided, but we can strongly suppress it. To this end, we also aim for objective \textit{(iii)} to penalize high mean photon numbers.

Finally, we state explicit criteria for our three objectives. To this end, we introduce the coefficient vector $\vec{c}\in\mathset{C}^L$ for decomposing the input (output) states into the $\{\ket{x_l}\}$ ($\{\ket{y_l}\}$) basis defined above. As a short-hand notation, we write $\ket{x(\vec{c})}=\sum_{l=1}^{L}c_l\ket{x_l}\in\mathcal{X}$ and $\ket{y(\vec{c})}=\sum_{l=1}^{L}c_l\ket{y_l}\in\mathcal{Y}$.

A reasonable figure of merit to quantify coherent errors is given by the mean overlap
\begin{equation}
	\quant{fidelity}[\hat{U}] :=
	\max_{\varphi\in\mathset{R}} \avg_{\substack{\text{$\vec{c}\in\mathset{C}^L$ \abbr{st}}\\\norm{\vec{c}}=1}} \Real{\Braket{y(\vec{c})}{\mconst{e}^{\mconst{i}\varphi}\hat{U}}{x(\vec{c})}}.
	\label{eq:def_mean_overlap}
\end{equation}
We use this to quantify objective \textit{(i)}, with $\hat{U}$ chosen according to \cref{eq:def_snap_and_displacment,eq:def_gate_sequence}, \abbr{ie} neglecting noise and gate imperfections. Property \textit{(ii)}, to aim for efficient sequences, simply means that $T$ should be small. Because SNAP gates take much longer than displacements, objective \textit{(iii)} can be achieved by minimizing $\sum_{t=1}^{T}\bar{n}_t$ where $\bar{n}_t$ denotes the mean photon number while executing $\hat{S}(\vec{\theta}_t)$, averaged over all input states. Since SNAP gates conserve the photon number, $\bar{n}_t$ can be computed as
\begin{equation}
	\bar{n}_t =
	\avg_{\substack{\text{$\vec{c}\in\mathset{C}^L$ \abbr{st}}\\\norm{\vec{c}}=1}} \Braket{x_t(\vec{c})}{\hat{n}}{x_t(\vec{c})}
	\label{eq:def_mean_photon_number}
\end{equation}
where
\begin{equation}
	\ket{x_t(\vec{c})}=\hat{D}(\alpha_t)\cdot\hat{S}(\vec{\theta}_{t-1})\cdot\hdots\cdot\hat{D}(\alpha_2)\cdot\hat{S}(\vec{\theta}_1)\cdot\hat{D}(\alpha_1)\ket{x(\vec{c})}
\end{equation}
represents the state immediately before applying $\hat{S}(\vec{\theta}_t)$.

\section{Results}
\label{sec:results}

Before we describe in \cref{sec:technique} how our technique works, we will present here results for a range of typical applications. Our intention behind this ordering is to first develop more intuition for the problem, and to show in advance some use cases which will help to understand certain design aspects for our technique.

\subsection{Operations on a 10-dimensional Fock space}
\label{sec:results:universal_control}

\begin{figure}[t!]
	\centering
	\includegraphics{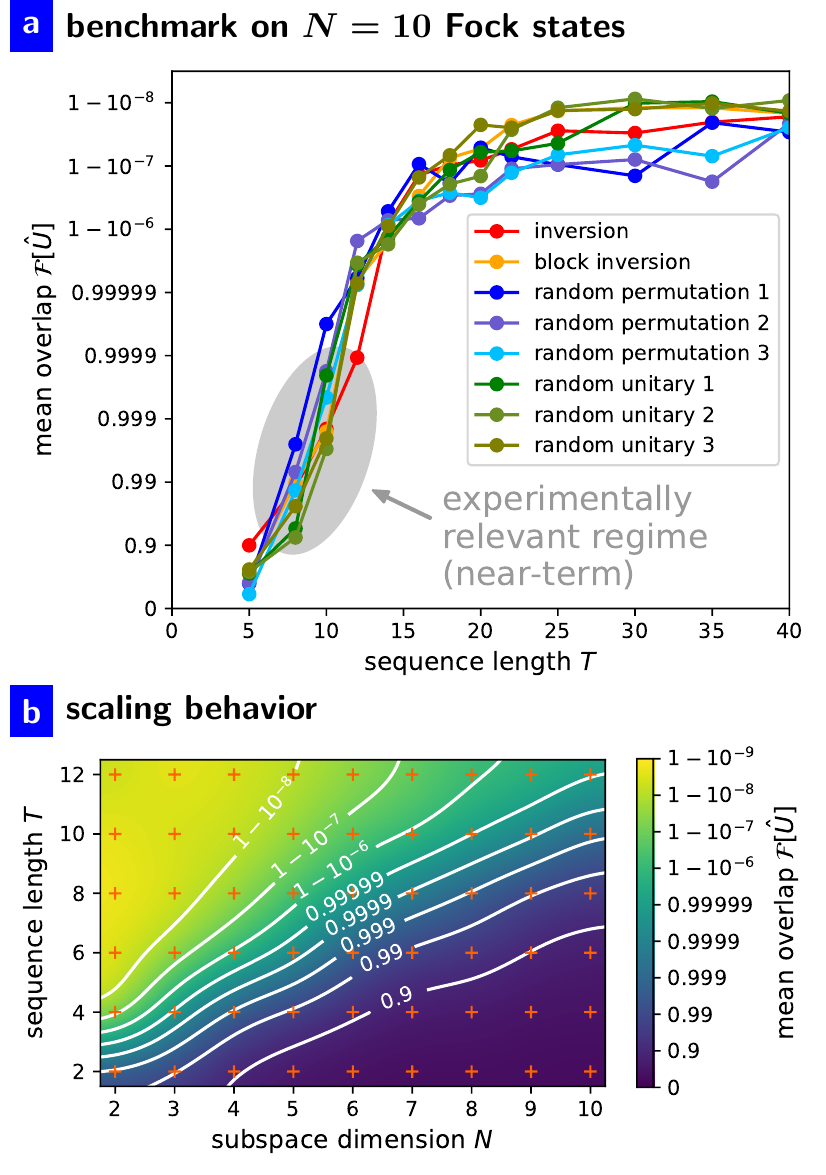}
	\caption{Results for operations on Fock subspaces.
		a) Benchmark for our technique, showing significantly reduced sequence lengths. The target operations are unitary transformations on the $N=10$ lowest Fock states. We apply our scheme to these target operations at different sequence lengths $T$, and plot the achieved mean overlap $\quant{fidelity}$ as a function of $T$. We see that longer sequences yield higher values for $\quant{fidelity}$, which is computed disregarding noise and gate imperfections. We highlight the regime with a good overall trade-off considering these effects. To improve visibility, the $y$ axis is scaled such that the error $1-\quant{fidelity}$ decreases logarithmically.
		b) Scaling behavior. The colors indicate how well sequences with $T$ SNAP gates can approximate target operations on the $N$ lowest Fock states. The isocurves roughly follow lines through the origin, which indicates that the sequence length for our method scales linearly in $N$. The values at the points marked with a cross are based on simulations (see \cref{sec:results:scaling_behavior}), and we use bicubic interpolation to extend this discrete grid of points to the entire plane.
	}
	\label{fig:results_unictrl}
\end{figure}

As a first benchmark for our technique, we consider transformations of the entire subspace spanned by the 10 lowest Fock states. To demonstrate generality, we choose eight target operations from different problem classes (\abbr{cmp} \cref{sec:appendix:examples:unictrl}). For each of them, we generate gate sequences of varying length. We expect that adding more gates, \abbr{ie} more control parameters, improves the ability to approximate the target operation. The achieved mean overlap for these gate sequences is plotted in \cref{fig:results_unictrl}a. Indeed, we find that the overlap steeply increases with the sequence length. In addition, we see that the same level of fidelity can be achieved with a comparable number of gates across the different examples.

These results constitute a significant improvement compared to the technique in \cite{krastanov2015snap_theory}, which here could only generate sequences with at least 100 SNAP gates. For comparable examples, this yielded a mean overlap in the order of $0.999$, which can be reached with \abbr{ca} 10 SNAP gates by our technique, \abbr{ie} with \abbr{ca} 10 times less gates!

In an experiment, one would choose a sequence length that minimizes the overall infidelity, including noise and gate imperfections. Because these effects grow with each new gate, the optimum is shifted towards shorter sequences. For an error probability per SNAP gate in a broad range between $\SI{0.01}{\percent}$ and $\SI{1}{\percent}$, the best trade-off is achieved around $T=8$ to $T=10$. This regime is highlighted as ``experimentally relevant'' in \cref{fig:results_unictrl}a.

We observe that the mean overlap saturates at a level between $1-10^{-7}$ and $1-10^{-9}$. Similar behavior will also occur in the following examples. Because we are not aware of any reason for this fidelity regime to be the ultimate limit, we assume that the saturation is a property of our numerical scheme. However, this question is of minor practical relevance: for a gate sequence with this precision, the other error sources like noise and gate imperfections will become the limiting factors for the overall fidelity, so further optimization of this contribution would effectively not help. The saturation would only become a bottleneck for an error probability per gate of less than $\SI{e-7}{}$.

Besides the sequence length and the fidelity, the third important property of a gate sequence is the photon number, which determines the rate of photon losses. To improve this aspect, we have equipped our technique with a mechanism to prefer low photon numbers during optimization (\abbr{cmp} \cref{sec:technique:finetuning}). In order to quantify its success, we consider $\sum_{t=1}^{T}\bar{n}_t$ where $\bar{n}_t$ is the mean photon number while executing the $t$-th SNAP gate (see \cref{eq:def_mean_photon_number}). For our examples, we find that $\frac{1}{T}\sum_{t=1}^{T}\bar{n}_t$ ranges from $4.51$ to $5.76$, and is less than $5.0$ in $\SI{80}{\percent}$ of the cases. For comparison, the minimum value for $\bar{n}_t$ to store $10$ states is $4.5$.

\subsection{Scaling behavior}
\label{sec:results:scaling_behavior}

In the introduction, we have argued that for fixed fidelity, the minimum sequence length, characterized by $T$, should scale linearly with the subspace dimension $N$. We now analyse the scaling behavior for the gate sequences generated by our technique.

To this end, we choose three random unitary transformations of the $N$ lowest Fock states as the target operations, for each value of $N$ between $2$ and $10$. The target operations for $N=10$ coincide with the random unitaries from \cref{fig:results_unictrl}a. We apply our technique to these target operations for all even sequence lengths $T$ between $2$ and $12$. Then, we compute their mean overlap $\quant{fidelity}$, and determine for all tuples $(N,T)$ an average value according to\footnote{The linear average gives a similar picture, but is more prone to random deviations.} $-\ln(1-\bar{\quant{fidelity}})=\langle-\ln(1-\quant{fidelity}_j)\rangle_j$. The result is plotted in \cref{fig:results_unictrl}b.

\Cref{fig:results_unictrl}b indicates that, at least for this class of operations, the sequence length to achieve similar overlap indeed grows linearly with $N$, since the isocurves roughly follow lines through the origin. This linear scaling behavior coincides with the expectation for the minimally required sequence length according to our parameter-counting argument. However, this does not yet mean that our sequences are optimum (or close-to-optimum), as there could be a difference in the constant of proportionality.

Based on our results, we can compute a rough estimate for the mean overlap as $\quant{fidelity}\approx1-\mconst{e}^{-6.49\cdot(T/N)^{1.91}}$ (\abbr{cmp} \cref{fig:suppl_scaling_behavior}). This relation is an empirical observation from the data points, but we do not have a model to predict this behavior.

\subsection{Applied problems}

\begin{figure*}[t!]
	\centering
	\includegraphics{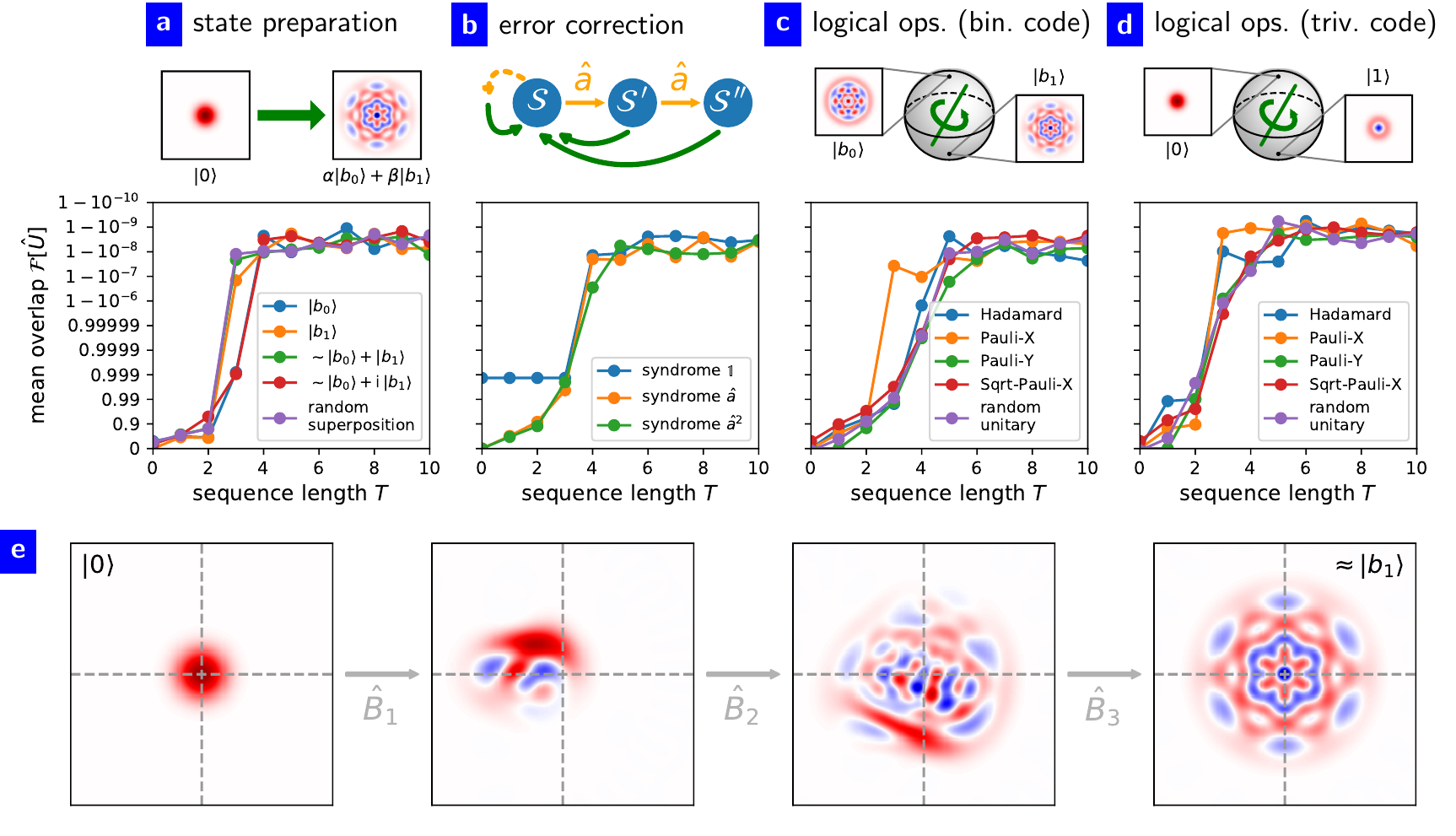}
	\caption{Results for experimentally relevant applications, showing that high fidelity for these operations can be achieved with a remarkably small number of gates.
		a) State preparation. A binomial code state $\alpha\ket{b_0}+\beta\ket{b_1}$ is generated from the vacuum state $\ket{0}$.
		b) Error correction for the binomial code $\alpha\ket{b_0}+\beta\ket{b_1}$. After a syndrome detection, the system needs to be transformed back into the code space $\mathcal{S}=\mathrm{span}(\ket{b_0},\ket{b_1})$; also the $\unitmat$ syndrome requires a small correction, see main text.
		c) Logical operations on the binomial code $\alpha\ket{b_0}+\beta\ket{b_1}$. The logical state, encoded in the coefficients $\alpha$ and $\beta$, is modified via a unitary transformation of $(\alpha,\beta)$. This operation must be performed without knowledge of the logical state.
		d) Logical operations on the trivial code $\alpha\ket{0}+\beta\ket{1}$. As in (c), but for a different encoding.
		The upper row in (a)-(d) illustrates the target operation, and the lower row plots the achieved fidelities (disregarding noise and gate imperfections) as a function of the sequence length.
		e) Evolution of the Wigner density for preparation of the binomial code state $\vert b_1\rangle$ (\abbr{cmp} a). The pictures display snapshots between different building blocks (\abbr{cmp} \cref{eq:def_building_blocks}). With every step, the structure of the state gains complexity, and also the volume in phase space is growing.
	}
	\label{fig:results_applications}
\end{figure*}

We will now consider several experimentally relevant problems for quantum information processing in cavities, and demonstrate how our technique can be beneficial for them.

In circuit QED, the quantum information is typically stored in the cavity. In the simplest case, we have a single logical qubit which is encoded in the two lowest Fock states:
\begin{equation}
	\ket{\psi_{\alpha\beta}^\mathrm{(triv)}} = \alpha \ket{0} + \beta \ket{1}.
	\label{eq:trivial_code}
\end{equation}
We refer to this as the trivial code. This code has the issue that it is very vulnerable to noise. To overcome this problem, several error-correction codes have been developed in the recent years. Among the most important codes for cavities are the binomial (or \textit{kitten}) codes\cite{michael2016kitten_code_theory,hu2019kitten_code_experiment}. In this work, we will specifically consider the binomial code $\ket{\psi_{\alpha\beta}^\mathrm{(bin)}}=\alpha\ket{b_0}+\beta\ket{b_1}$ with
\begin{align}
	\ket{b_0} & = \frac{\ket{0}+\sqrt{3}\ket{6}}{2} &
	\ket{b_1} & = \frac{\sqrt{3}\ket{3}+\ket{9}}{2}
	\label{eq:kitten_code}
\end{align}
which protects against single- and double-photon loss. In principle, this code could also tolerate a single dephasing error, but this noise channel plays only a negligible role if the cavity evolves freely.

The first application we consider is how to prepare the cavity in one of these binomial code states $\ket{\psi_{\alpha\beta}^\mathrm{(bin)}}$, starting from the vacuum state $\ket{0}$. This is an example where the transformation of only one state matters, so we have a one-dimensional logical subspace ($L=1$). The achieved mean overlap for preparing several superpositions of $\ket{0}$ and $\ket{1}$ is shown in \cref{fig:results_applications}a, as a function of the sequence length. We see that high fidelity can be reached within only $3$ or $4$ SNAP gates!

As a second application, we consider recovery operations. The main benefit of an error-correction code is to protect the quantum information against certain noise events, in the case of the binomial code from \cref{eq:kitten_code} against up to two photon losses. To avoid that over time more of these events accumulate than which can be corrected for, the error syndrome (here: photon number modulo $3$) has to be monitored frequently. If such a syndrome detection reveals a single- or double-photon loss, the corresponding number of photons needs to be repumped to return into the original, robust code. Even if no photon loss is detected, a small correction needs to be applied to counter the effect of amplitude decay, \abbr{ie} that the amplitudes for higher Fock states decay faster (\abbr{cmp} \cref{sec:appendix:examples:applied}). Hence, an active recovery operation is required in all three cases. This operation must not depend on the logical state, which is unknown and not allowed to be measured, so a two-dimensional subspace needs to be transformed ($L=2$). Again, our technique can be used to generate efficient gate sequences, as shown in \cref{fig:results_applications}b: $4$ SNAP gates are sufficient to perform these operations with high fidelity.

Another related task is to perform logical operations, \abbr{ie} to effect a unitary transformation of the coefficients $\alpha$ and $\beta$, here on the binomial code $\alpha\ket{b_0}+\beta\ket{b_1}$. As for the previous problem, the operation must not depend on the incoming logical state, so again the logical subspace is two-dimensional ($L=2$). A special case are rotations around the logical $z$ axis which can be realized with a single SNAP gate. All other operations, however, require more complex control schemes. \Cref{fig:results_applications}c shows the results for five selected examples. We reach good fidelity with $4$, in one case $3$, SNAP gates.

We observe that all these operations can be realized with between $3$ and $4$ SNAP gates using our technique. This value is exceeded considerably by the \textit{minimum} sequence length for the method from \cite{krastanov2015snap_theory} in almost every case, as shown in \cref{sec:appendix:examples:comparison}. The largest improvement can be obtained for the error correction examples (\cref{fig:results_applications}b), which are also the most time-critical operations. Here, the minimum number of SNAP gates for the technique in \cite{krastanov2015snap_theory} would be $16$ for the syndrome $\unitmat$, and $49$ for the syndroms $\hat{a}$ and $\hat{a}^2$.

In addition, the sequences here are significantly shorter than what would be expected from \cref{sec:results:scaling_behavior}, taking into account their respective number of involved Fock states. This is possible because here only one or two states within this subspace are actually relevant. In these cases, we benefit from our scheme to focus solely on the action on the input states (\abbr{cmp} \cref{sec:technique:properties}).

The examples so far show how our technique can be used to efficiently implement the relevant operations for one binomial code. However, similar problems occur also under other circumstances. Recovery operations as in \cref{fig:results_applications}b are needed for every cavity code that requires active error correction. Logical operations as in \cref{fig:results_applications}c occur for every encoding, whether error-corrected or not. State preparation as in \cref{fig:results_applications}a has many applications even outside of quantum information processing. Our technique is also useful for those cases, and by no means restricted to the binomial code.

As an example, we consider one more application, namely logical operations on the trivial code as in \cref{eq:trivial_code}. We consider the same logical operations as in \cref{fig:results_applications}c, but on the code $\alpha\ket{0}+\beta\ket{1}$ instead of $\alpha\ket{b_0}+\beta\ket{b_1}$. The results, as plotted in \cref{fig:results_applications}d, indicate that sequences with $3$ SNAP gates can perform these operations with sufficiently high fidelity.

\section{Technique}
\label{sec:technique}

\begin{figure}[b!]
	\centering
	\includegraphics{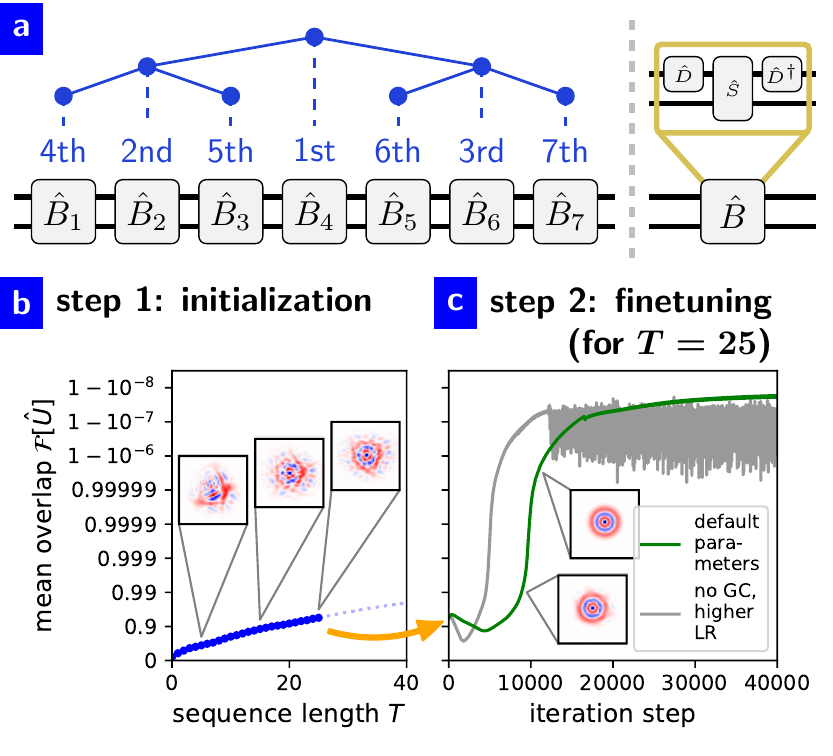}
	\caption{Construction process.
		a) Hierarchical insertion strategy for the initialization, which lets the fidelity improve progressively during this whole process. The order in which the building blocks $\hat{B}_t=\hat{D}^\dagger(\alpha_t)\hat{S}(\vec{\theta}_t)\hat{D}(\alpha_t)$ are inserted corresponds to the breadth-first traversal of a binary tree (here until sequence length $T=7$).
		b) Progress during initialization, from an empty sequence towards a coarse approximation of the target operation. The curve shows how the mean overlap $\quant{fidelity}$ improves by inserting new gates into the sequence.
		c) Progress during finetuning, showing the attainment of high fidelity. After stopping initialization, here at sequence length $T=25$, we further improve the fidelity by co-optimizing the control parameters. With our default hyperparameters (green curve), the fidelity grows steeply within the first $15000$ iterations (after a first phase where mainly the photon number is optimized), surpassing a value of $1-10^{-7}$. Afterwards, the improvement gets much slower. For comparison, the gray curve shows the progress without gradient clipping and for a higher learning rate. While this parameter configuration learns faster in the beginning, it becomes unstable in the high-fidelity regime. However, post-selecting good intermediate values leads to comparable overall results as with the default parameters. The exact behavior can vary from case to case.
		The insets in (b) and (c) show the final state that should ideally be mapped to the state $\ket{3}$. The target operation is ``random unitary 3'' from \cref{fig:results_unictrl}a.
	}
	\label{fig:training}
\end{figure}

After presenting the results, we now turn to a description of our technique. It consists of two parts: an initialization step, followed by a finetuning step. Whereas this basic concept has already been used in \cite{krastanov2015snap_theory}, we have fundamentally redesigned the approaches towards both of these steps. These changes enable us to arrive at significantly better results.

In the initialization step, we use a heuristic strategy to generate a gate sequence which roughly approximates the target operation. The goal is not yet to achieve high fidelity, as eventually required in the experiment, and usually we obtain just a coarse approximation (\abbr{cmp} \cref{fig:training}b). Instead, we delegate the responsibility for optimizing the fidelity mainly into the finetuning step, and are here content with providing a reasonable starting point for this. In contrast to the fidelity, the length of the sequence will not be changed during finetuning. Our new initialization scheme can construct suitable sequences of arbitrary length, in particular also at relatively small numbers of gates. This is a central advantage compared to its counterpart in \cite{krastanov2015snap_theory} (which is based on $\mathset{SO2}$ rotations).

In the finetuning step, we use gradient descent to co-optimize the control parameters of the gate sequence. This is necessary to reach the high-fidelity regime (\abbr{cmp} \cref{fig:training}c) which is needed for application in an experiment. We require from this process that it can be launched successfully from the starting point provided by the initialization step, for which the mean overlap is typically between $0.75$ and $0.995$.

During both steps, we will work with building blocks of the form
\begin{equation}
	\hat{B}(\alpha,\vec{\theta}) = \hat{D}^\dagger(\alpha) \hat{S}(\vec{\theta}) \hat{D}(\alpha)
	\label{eq:def_building_blocks}
\end{equation}
where $\hat{D}(\alpha)$ denotes a cavity displacement and $\hat{S}(\vec{\theta})$ a SNAP gate. The direct output of our technique are sequences of these building blocks. To implement them on hardware, we can easily reparameterize them in terms of SNAP gates and displacements (\abbr{cmp} \cref{sec:appendix:gate_seq:building_blocks}). This equivalence also guarantees these building blocks to be computationally universal.

\subsection{Initialization}
\label{sec:technique:initialization}

As motivated above, the goal for the initialization step is rather to approximate the target operation than to aim directly for very high fidelity. The more important aspect is the flexibility regarding the length of the sequence. In the following, we describe our initialization scheme which starts from an empty sequence and expands it successively with the building blocks from \cref{eq:def_building_blocks}.

We first consider, for simplicity, the special case of a gate sequence with only one building block $\hat{B}(\alpha,\vec{\theta})$. In this situation, we can directly compute the parameters $\alpha$ and $\vec{\theta}$ which maximize the mean overlap $\quant{fidelity}[\hat{B}(\alpha,\vec{\theta})]$ as
\begin{align}
	\alpha & = \argmax_{\tilde{\alpha}} \Big[\sum_n\abs{g_n(\tilde{\alpha})}\Big] &
	\theta^{(n)} & = \arg(g_n(\alpha))
	\label{eq:initialization}
\end{align}
where $g_n(\alpha)=\Braket{n}{\hat{D}(\alpha)\hat{V}\hat{D}^\dagger(\alpha)}{n}$ and $\hat{V}$ encodes the target operation (\abbr{cmp} \cref{eq:definition_target_operation}). For the derivation, see \cref{sec:appendix:initialization}.

With a slight modification, we can use this scheme also to successively extend an existing gate sequence. In each step, one building block $\hat{B}_t$ is inserted between the already fixed segments $\hat{B}_T,\hdots,\hat{B}_{t+1}$ and $\hat{B}_{t-1},\hdots,\hat{B}_1$. We choose $\hat{B}_t$ such that it greedily maximizes the mean overlap $\quant{fidelity}[\hat{B}_T\cdot\hdots\cdot\hat{B}_1]$ of the extended sequence. Afterwards, also this $\hat{B}_t$ remains fixed for the rest of this procedure, and we continue with inserting the next building block (for a new subdivision into the two segments) until the desired sequence length is reached.

The greedy optimization of $\hat{B}_t=\hat{B}(\alpha_t,\vec{\theta}_t)$ is an equivalent problem to matching the input states $\hat{B}_{t-1}\cdot\hdots\cdot\hat{B}_1\ket{x(\vec{c})}$ to the output states $\hat{B}_{t+1}^\dagger\cdot\hdots\cdot\hat{B}_T^\dagger\ket{y(\vec{c})}$ with a single building block. Hence, by replacing $\hat{V}$ with $\hat{B}_{t+1}^\dagger\cdot\hdots\cdot\hat{B}_T^\dagger\cdot\hat{V}\cdot\hat{B}_1^\dagger\cdot\hdots\cdot\hat{B}_{t-1}^\dagger$, we can reuse our solution for this situation: the parameters $\alpha_t$ and $\vec{\theta}_t$ can be determined according to \cref{eq:initialization}, where
\begin{equation}
	g_n(\alpha) = \Braket[\big]{n}{\hat{D}(\alpha)\cdot\hat{B}_{t+1}^\dagger\cdot\hdots\cdot\hat{B}_T^\dagger\cdot\hat{V}\cdot\hat{B}_1^\dagger\cdot\hdots\cdot\hat{B}_{t-1}^\dagger\cdot\hat{D}^\dagger(\alpha)}{n}.
\end{equation}
This choice guarantees to never reduce the fidelity because there is always the trivial option $\hat{B}_t=\unitmat$, so the old value for the fidelity is a lower bound for the new one. In practice, we find that typically the fidelity gradually increases with every new building block (\abbr{cmp} \cref{fig:training}b).

One degree of freedom in this procedure is the order in which the building blocks are inserted. We find that for a linear arrangement, \abbr{ie} building the sequence either from start to end or from end to start, the improvement of the fidelity would stagnate after a few steps because only operations close to identity will be inserted. Besides its lower fidelity, such a sequence has the issue that it does not at all make efficient use of its expressive power, which impedes subsequent finetuning. This problem can be circumvented by inserting the building blocks hierarchically between the previously added ones (\abbr{cmp} \cref{fig:training}a). We follow this strategy during the initialization for all results shown in \cref{sec:results}.

For one example, the recovery of the $\unitmat$ syndrome for a binomial code (\abbr{cmp} \cref{fig:results_applications}b), we have to slightly deviate from the scheme as described so far. Because all $g_n(\alpha)$ take real values here, we would run into the issue that only identity operations would be inserted. We break this blockade by starting the initialization with one randomly chosen building block and then continue as usual.

For this strategy of expanding sequences stepwise, it is very advantageous to have the building blocks parameterized as $\hat{D}^\dagger(\alpha)\hat{S}(\vec{\theta})\hat{D}(\alpha)$. In particular, such building blocks do not generate a net displacement, in contrast to other parameterizations like $\hat{S}(\vec{\theta})\hat{D}(\alpha)$, and therefore already isolated building blocks can be useful for the important case of displacement-free target operations. The importance of choosing a good parameterization has been observed in other platforms before \cite{zeuch2020efficient_sequences}.

\subsection{Finetuning}
\label{sec:technique:finetuning}

After the initialization, we perform an additional finetuning step where all the control parameters in the gate sequence (displacements $\alpha_t$ and SNAP angles $\vec{\theta}_t$) are co-optimized. We accomplish this using gradient descent, using the cost function
\begin{equation}
	C(\alpha_1,\vec{\theta}_1,\hdots,\alpha_T,\vec{\theta}_T)
	=
	\ln\big(1-\quant{fidelity}[\hat{B}_T\cdot\hdots\cdot\hat{B}_1]\big) + \lambda \sum_{t=1}^{T} \frac{\bar{n}_t+\bar{n}_t'}{2}
	\label{eq:finetuning_cost_function}
\end{equation}
where $\quant{fidelity}$ is the mean overlap from \cref{eq:def_mean_overlap}, $\bar{n}_t$ is the mean photon number while executing $\hat{S}(\vec{\theta}_t)$ as defined in \cref{eq:def_mean_photon_number}, and $\bar{n}_t'$ is the corresponding quantity for the inverse sequence starting from the desired output states $\ket{y(\vec{c})}$ (see \cref{sec:appendix:objectives:summary}).

The term $\ln(1-\quant{fidelity}[\hat{B}_T\cdot\hdots\cdot\hat{B}_1] )$ is responsible for matching the target operation. We use this logarithmic expression instead of $-\quant{fidelity}[\hat{B}_T\cdot\hdots\cdot\hat{B}_1]$ to avoid that the gradient diminishes in the high-fidelity regime, \abbr{ie} when $\quant{fidelity}[\hat{B}_T\cdot\hdots\cdot\hat{B}_1]$ gets close to $1$. In practice, this significantly accelerates the training process.

The term $\sum_{t=1}^{T}\frac{\bar{n}_t+\bar{n}_t'}{2}$ penalizes larger photon numbers in the cavity, which would amplify the rate of photon losses. Without this contribution to the cost function, or if its coefficient $\lambda$ was chosen too small, higher Fock states would get populated stronger than necessary during the gate sequence. Increasing $\lambda$ helps to reduce this effect, and often even generates sequences with a higher mean overlap $\quant{fidelity}$. However, for $\lambda$ too large, reducing the photon number would become at some point more favorable than matching the target operation. We have always found good intermediate values for $\lambda$ with high fidelity and low photon number, but this range can vary from example to example (for us, between $\lambda=0.16$ and $\lambda=2.4$, \abbr{cmp} \cref{sec:appendix:simulations:hyperparams}).

We compute the parameter updates from the gradient according to the Adam method \cite{kingma2014adam}. In addition, we use gradient clipping to stabilize the training process (\abbr{cmp} \cref{fig:training}c). These are two well-established concepts from the field of machine learning, and they help us to reach convergence with a significantly reduced number of iterations. To keep the runtime of the simulations moderate, we need in addition an efficient numerical scheme to compute the gradient (\abbr{cmp} \cref{sec:appendix:finetuning:overlap}).

\subsection{Discussion}
\label{sec:technique:properties}

Our scheme is carefully designed such that it focuses only on the relevant aspect of the operation, namely on the mapping between the specified input and output states, and does on purpose not consider the action on any other state. This goes further than merely ignoring the action on Fock states which have no overlap with any input or output state: The previous scheme from \cite{krastanov2015snap_theory} requires to fix explicitly a unitary on the full subspace of involved Fock states, even if the aim was only to transform a single input state into one desired output state. In contrast, our approach keeps the number of conditions minimal, which allows to improve the behavior on the relevant states. On the technical side, this means that we must not use an explicit extension of $\hat{V}$ (from \cref{eq:definition_target_operation}) to a unitary operator $\hat{V}_\mathrm{ext}$, \abbr{eg} for a figure of merit like $\abs{\tr[\hat{V}_\mathrm{ext}^\dagger\hat{U}]}$. Instead, both during initialization and finetuning, we maximize $\quant{fidelity}[\hat{U}]\sim\abs{\tr[\hat{V}^\dagger\hat{U}]}$ (\abbr{cmp} \cref{sec:appendix:objectives:summary}) which is directly based on $\hat{V}$. A similar strategy is followed by \cite{heeres2017grape} in their optimal-control pulse engineering.

Our technique involves only a few hyperparameters. The only one which is changed between the different examples is $\lambda$, the coefficient for the photon number cost (see \cref{sec:technique:finetuning}). Depending on the application, the range of good values can span a factor between 2 and more than 100. For all other hyperparameters, the same values are used for all results in \cref{sec:results}. All hyperparameter values are listed in \cref{sec:appendix:simulations:hyperparams}.

Even though we do not employ neural networks, we can still benefit a lot from machine-learning techniques: our finetuning approach makes use of the Adam optimizer, gradient clipping, and a backpropagation-like scheme to compute the gradient. In this sense, our work is related to previous applications of machine learning for problems related to quantum computing \cite{torlai2017neural_decoder,melnikov2018active,foesel2018rl_for_qec}, and especially to quantum control \cite{bukov2018rl_quantum_ctrl,august2018blackbox_quantum_ctrl,niu2019universal_quantum_ctrl,porotti2019quantum_transport_rl}. However, there is still a big difference in the concrete methodology, as these works use neural networks and/or reinforcement learning. Another aspect that sets us apart (from the mentioned works on quantum control) is the considered control problem: their goal is to engineer pulses for single gates, whereas here we focus on building sequences by combining multiple gates.

\section{Conclusion}

In this work, we have presented a practical scheme to construct efficient sequences of SNAP gates and displacements for given target operations. We have demonstrated that our scheme works robustly for a broad range of applications, including tasks of direct experimental relevance. Moreover, our scheme is directly applicable for experimentalists: all necessary gates have already been demonstrated, and our technique outputs explicitly their relevant control parameters.

Like other gradient-descent techniques, our technique likely does not find the global optimum, and in principle we cannot even be sure that it finds close-to-optimal solutions. Nevertheless, in practice, our scheme can generate much shorter gate sequences compared to previously known techniques, with higher fidelity. The consequence of this improvement is that realizing universal control over cavities with SNAP gates becomes experimentally feasible: As we have demonstrated above, we can obtain accelerations by one order of magnitude, which turns out to take us into the regime where gate times are sufficiently short given the coherence of current devices. In this way, our work also reveals the power and flexibility of the SNAP gate when used properly in combination with displacements.

In our simulations, we have not considered directly the effect of gate imperfections and noise. However, our technique 
is able to considerably reduce the sequence length; this already mitigates their influence very effectively. A more rigorous approach is to include these effects into the simulation, which is subject to future work.

Our work is compatible with the error-transparency scheme in \cite{reinhold2019snap_error_transparency,ma2019path_independent_gates}, from which we can gain robustness against transmon decay to first order. Accelerating the SNAP gates while correcting for coherent errors, \abbr{eg} using a Magnus-based approach \cite{roque2020magnus}, could further improve the overall fidelity.

We have applied our technique explicitly for sequences of SNAP gates and displacements, due to their direct relevance for circuit QED experiments. However, variations of our technique may be used to optimize control schemes for other platforms, too. Moreover, our combination of initialization and finetuning is a more general concept which has possible applications even outside of quantum control, namely for any continous control problem on discrete time steps without feedback.

\textit{Data and code availability} The data and the code that support the plots within this paper will be made publicly available. Other findings of this study are available from the corresponding author on request.

\textit{Acknowledgements} We thank Hugo Ribeiro, Thales Figueiredo Roque, Wenlong Ma and Christopher Wang for fruitful discussions. T.\,F.\ thanks Yale Quantum Institute for hosting during his research stay. L.\,J.\ acknowledges supports from the ARL-CDQI (W911NF-15-2-0067), ARO (W911NF-18-1-0020, W911NF-18-1-0212), ARO MURI (W911NF-16-1-0349), AFOSR MURI (FA9550-15-1-0015, FA9550-19-1-0399), DOE (DE-SC0019406), NSF (EFMA-1640959, OMA-1936118), and the Packard Foundation (2013-39273).

\bibliography{references}{}

\begin{thebibliography}{10}

\bibitem{nielsen2011book}
Michael~A. Nielsen and Isaac~L. Chuang.
\newblock {\em Quantum {Computation} and {Quantum} {Information}: 10th
  {Anniversary} {Edition}}.
\newblock Cambridge University Press, Cambridge ; New York, anniversary edition
  edition, January 2011.

\bibitem{ladd2010review_qc}
Thaddeus~D Ladd, Fedor Jelezko, Raymond Laflamme, Yasunobu Nakamura,
  Christopher Monroe, and Jeremy~Lloyd O’Brien.
\newblock Quantum computers.
\newblock {\em Nature}, 464(7285):45--53, 2010.

\bibitem{schoelkopf2008review_superconducting_qubits}
RJ~Schoelkopf and SM~Girvin.
\newblock Wiring up quantum systems.
\newblock {\em Nature}, 451(7179):664--669, 2008.

\bibitem{devoret2013review_superconducting_qubits}
Michel~H Devoret and Robert~J Schoelkopf.
\newblock Superconducting circuits for quantum information: an outlook.
\newblock {\em Science}, 339(6124):1169--1174, 2013.

\bibitem{wendin2017superconducting_qubits}
G~Wendin.
\newblock Quantum information processing with superconducting circuits: a
  review.
\newblock {\em Reports on Progress in Physics}, 80(10):106001, 2017.

\bibitem{ofek2016cat_code_experiment}
Nissim Ofek, Andrei Petrenko, Reinier Heeres, Philip Reinhold, Zaki Leghtas,
  Brian Vlastakis, Yehan Liu, Luigi Frunzio, SM~Girvin, L~Jiang, et~al.
\newblock Extending the lifetime of a quantum bit with error correction in
  superconducting circuits.
\newblock {\em Nature}, 536(7617):441, 2016.

\bibitem{axline2018on_demand}
Christopher~J Axline, Luke~D Burkhart, Wolfgang Pfaff, Mengzhen Zhang, Kevin
  Chou, Philippe Campagne-Ibarcq, Philip Reinhold, Luigi Frunzio, SM~Girvin,
  Liang Jiang, et~al.
\newblock On-demand quantum state transfer and entanglement between remote
  microwave cavity memories.
\newblock {\em Nature Physics}, 14(7):705, 2018.

\bibitem{chou2018deterministic_teleportation}
Kevin~S Chou, Jacob~Z Blumoff, Christopher~S Wang, Philip~C Reinhold,
  Christopher~J Axline, Yvonne~Y Gao, Luigi Frunzio, MH~Devoret, Liang Jiang,
  and RJ~Schoelkopf.
\newblock Deterministic teleportation of a quantum gate between two logical
  qubits.
\newblock {\em Nature}, 561(7723):368, 2018.

\bibitem{heeres2017grape}
Reinier~W Heeres, Philip Reinhold, Nissim Ofek, Luigi Frunzio, Liang Jiang,
  Michel~H Devoret, and Robert~J Schoelkopf.
\newblock Implementing a universal gate set on a logical qubit encoded in an
  oscillator.
\newblock {\em Nature Communications}, 8(1):94, 2017.

\bibitem{khaneja2005grape}
Navin Khaneja, Timo Reiss, Cindie Kehlet, Thomas Schulte-Herbr{\"u}ggen, and
  Steffen~J Glaser.
\newblock Optimal control of coupled spin dynamics: design of nmr pulse
  sequences by gradient ascent algorithms.
\newblock {\em Journal of Magnetic Resonance}, 172(2):296--305, 2005.

\bibitem{heeres2015snap_experiment}
Reinier~W Heeres, Brian Vlastakis, Eric Holland, Stefan Krastanov, Victor~V
  Albert, Luigi Frunzio, Liang Jiang, and Robert~J Schoelkopf.
\newblock Cavity state manipulation using photon-number selective phase gates.
\newblock {\em Physical Review Letters}, 115(13):137002, 2015.

\bibitem{krastanov2015snap_theory}
Stefan Krastanov, Victor~V Albert, Chao Shen, Chang-Ling Zou, Reinier~W Heeres,
  Brian Vlastakis, Robert~J Schoelkopf, and Liang Jiang.
\newblock Universal control of an oscillator with dispersive coupling to a
  qubit.
\newblock {\em Physical Review A}, 92(4):040303, 2015.

\bibitem{reinhold2019snap_error_transparency}
Philip Reinhold, Serge Rosenblum, Wen-Long Ma, Luigi Frunzio, Liang Jiang, and
  Robert~J Schoelkopf.
\newblock Error-corrected gates on an encoded qubit.
\newblock {\em arXiv preprint arXiv:1907.12327}, 2019.

\bibitem{ma2019path_independent_gates}
Wen-Long Ma, Mengzhen Zhang, Yat Wong, Kyungjoo Noh, Serge Rosenblum, Philip
  Reinhold, Robert~J Schoelkopf, and Liang Jiang.
\newblock Path-independent quantum gates with noisy ancilla.
\newblock {\em arXiv preprint arXiv:1911.12240}, 2019.

\bibitem{michael2016kitten_code_theory}
Marios~H Michael, Matti Silveri, RT~Brierley, Victor~V Albert, Juha Salmilehto,
  Liang Jiang, and Steven~M Girvin.
\newblock New class of quantum error-correcting codes for a bosonic mode.
\newblock {\em Physical Review X}, 6(3):031006, 2016.

\bibitem{hu2019kitten_code_experiment}
L~Hu, Y~Ma, W~Cai, X~Mu, Y~Xu, W~Wang, Y~Wu, H~Wang, YP~Song, C-L Zou, et~al.
\newblock Quantum error correction and universal gate set operation on a
  binomial bosonic logical qubit.
\newblock {\em Nature Physics}, 15(5):503, 2019.

\bibitem{zeuch2020efficient_sequences}
Daniel Zeuch and NE~Bonesteel.
\newblock Efficient two-qubit pulse sequences beyond cnot.
\newblock {\em arXiv preprint arXiv:2001.09341}, 2020.

\bibitem{kingma2014adam}
Diederik~P Kingma and Jimmy Ba.
\newblock Adam: A method for stochastic optimization.
\newblock {\em arXiv preprint arXiv:1412.6980}, 2014.

\bibitem{torlai2017neural_decoder}
Giacomo Torlai and Roger~G Melko.
\newblock Neural decoder for topological codes.
\newblock {\em Physical review letters}, 119(3):030501, 2017.

\bibitem{melnikov2018active}
Alexey~A Melnikov, Hendrik~Poulsen Nautrup, Mario Krenn, Vedran Dunjko, Markus
  Tiersch, Anton Zeilinger, and Hans~J Briegel.
\newblock Active learning machine learns to create new quantum experiments.
\newblock {\em Proceedings of the National Academy of Sciences},
  115(6):1221--1226, 2018.

\bibitem{foesel2018rl_for_qec}
Thomas F{\"o}sel, Petru Tighineanu, Talitha Weiss, and Florian Marquardt.
\newblock Reinforcement learning with neural networks for quantum feedback.
\newblock {\em Physical Review X}, 8(3):031084, 2018.

\bibitem{bukov2018rl_quantum_ctrl}
Marin Bukov, Alexandre~GR Day, Dries Sels, Phillip Weinberg, Anatoli
  Polkovnikov, and Pankaj Mehta.
\newblock Reinforcement learning in different phases of quantum control.
\newblock {\em Physical Review X}, 8(3):031086, 2018.

\bibitem{august2018blackbox_quantum_ctrl}
Moritz August and Jos{\'e}~Miguel Hern{\'a}ndez-Lobato.
\newblock Taking gradients through experiments: Lstms and memory proximal
  policy optimization for black-box quantum control.
\newblock In {\em International Conference on High Performance Computing},
  pages 591--613. Springer, 2018.

\bibitem{niu2019universal_quantum_ctrl}
Murphy~Yuezhen Niu, Sergio Boixo, Vadim~N Smelyanskiy, and Hartmut Neven.
\newblock Universal quantum control through deep reinforcement learning.
\newblock {\em npj Quantum Information}, 5(1):1--8, 2019.

\bibitem{porotti2019quantum_transport_rl}
Riccardo Porotti, Dario Tamascelli, Marcello Restelli, and Enrico Prati.
\newblock Coherent transport of quantum states by deep reinforcement learning.
\newblock {\em Communications Physics}, 2(1):1--9, 2019.

\bibitem{roque2020magnus}
Thales~Figueiredo Roque, Aashish~A Clerk, and Hugo Ribeiro.
\newblock Engineering fast high-fidelity quantum operations with constrained
  interactions.
\newblock {\em arXiv preprint arXiv:2003.12096}, 2020.

\end{thebibliography}
\bibliographystyle{unsrt}

\onecolumngrid
\newpage
\appendix

\section{Gate sequences}

\subsection{Ansatz}
\label{sec:appendix:gate_seq:ansatz}

In order to motivate the ansatz in \cref{eq:def_gate_sequence}, we start with the observation that two (or multiple) subsequent SNAP gates can be contracted into a single one, and the same statement holds also for displacements:
\begin{align*}
	\hat{S}(\vec{\theta}_2) \cdot \hat{S}(\vec{\theta}_1) & = \hat{S}(\vec{\theta}_1+\vec{\theta}_2) &
	\hat{D}(\alpha_2) \cdot \hat{D}(\alpha_1) & = \hat{D}(\alpha_1+\alpha_2).
\end{align*}
Hence, the most efficient sequences are those which always alternate between these two gate types:
\begin{equation*}
	\hat{U} = \hdots \cdot \hat{S}(\vec{\theta}_{t+1}) \cdot \hat{D}(\alpha_{t+1}) \cdot \hat{S}(\vec{\theta}_t) \cdot \hat{D}(\alpha_t) \cdot \hdots
\end{equation*}

The sequence structure is not yet characterized uniquely, as we can still decide whether to start, and also to end, such a sequence either with a SNAP gate or with a displacement. Because both the execution time and the gate errors for displacements are negligable compared to the corresponding values for SNAP gates, it makes sense to both start and end with a displacement, as these operations add a new degree of freedom to the sequence at very low costs:
\begin{equation*}
	\hat{U} = \hat{D}(\alpha_{T+1}) \cdot \hat{S}(\vec{\theta}_T) \cdot \hat{D}(\alpha_T) \cdot \hdots \cdot \hat{S}(\vec{\theta}_2) \cdot \hat{D}(\alpha_2) \cdot \hat{S}(\vec{\theta}_1) \cdot \hat{D}(\alpha_1)
\end{equation*}

The control parameters $\alpha_t$ can be restricted to real values because any complex phase can be absorbed into the neighboring SNAP gates, as implied by the relation $\hat{D}(\mconst{e}^{\mconst{i}\varphi}\alpha)=\hat{S}(\vec{\theta}_\mathrm{rot}(\varphi))\cdot\hat{D}(\alpha)\cdot\hat{S}(-\vec{\theta}_\mathrm{rot}(\varphi))$ with $\theta_\mathrm{rot}^{(n)}(\varphi)=n\varphi$. This argument applies to all displacement gates except for the first and the last one. In total, the full gate sequence loses only a single degree of freedom by the restriction to real $\alpha_t$.

\subsection{Reparameterization using building blocks}
\label{sec:appendix:gate_seq:building_blocks}

Here, we justify the usage of the building blocks $\hat{B}(\alpha,\vec{\theta})=\hat{D}^\dagger(\alpha)\hat{S}(\vec{\theta})\hat{D}(\alpha)$ as defined in \cref{eq:def_building_blocks}. These building blocks have several practical advantages over working directly with the native hardware gates $\hat{S}(\vec{\theta})$ and $\hat{D}(\alpha)$ during the construction of the gate sequence.

We can rewrite the gate sequence from \cref{eq:def_gate_sequence} as
\begin{align*}
	& \hat{D}(\alpha_{T+1}) \cdot \hat{S}(\vec{\theta}_T) \cdot \hat{D}(\alpha_T) \cdot \hdots \cdot \hat{S}(\vec{\theta}_2) \cdot \hat{D}(\alpha_2) \cdot \hat{S}(\vec{\theta}_1) \cdot \hat{D}(\alpha_1) = \\
	& = \hat{D}(\alpha_{T+1}') \cdot \hat{B}(\alpha_T', \vec{\theta}_T) \cdot \hdots \cdot \hat{B}(\alpha_2', \vec{\theta}_2) \cdot \hat{B}(\alpha_1', \vec{\theta}_1)
\end{align*}
with $\alpha_t'=\sum_{s=1}^{t}\alpha_s$.

A displacement operation $\hat{D}(\alpha)$ can be composed using two building blocks as
\begin{equation*}
	\hat{D}(\alpha) =
	\hat{B}(-{\textstyle\frac{\alpha-\delta\alpha}{4}},\vec{\theta}_\mathrm{rot}(\pi))
	\cdot
	\hat{B}({\textstyle\frac{\alpha+\delta\alpha}{4}},\vec{\theta}_\mathrm{rot}(\pi))
\end{equation*}
for $\vec{\theta}_\mathrm{rot}^{(n)}(\pi)=n\mconst{pi}$ and any $\delta\alpha\in\mathset{C}$ (set to $0$ in the next equation). This relation can be verified by expanding the right-hand side and using $\hat{S}(\vec{\theta}_\mathrm{rot}(\pi))\hat{D}(\alpha)\hat{S}(\vec{\theta}_\mathrm{rot}(\pi))=\hat{D}(-\alpha)$. This consideration guarantees that any sequence of SNAP gates and displacements can be exactly reparameterized with these building blocks:
\begin{align*}
	& \hat{D}(\alpha_{T+1}) \cdot \hat{S}(\vec{\theta}_T) \cdot \hat{D}(\alpha_T) \cdot \hdots \cdot \hat{S}(\vec{\theta}_2) \cdot \hat{D}(\alpha_2) \cdot \hat{S}(\vec{\theta}_1) \cdot \hat{D}(\alpha_1) = \\
	& = \hat{B}(-\alpha_{T+1}'/4,\vec{\theta}_\mathrm{rot}(\pi)) \cdot \hat{B}(\alpha_{T+1}'/4,\vec{\theta}_\mathrm{rot}(\pi)) \cdot \hat{B}(\alpha_T', \vec{\theta}_T) \cdot \hdots \cdot \hat{B}(\alpha_2', \vec{\theta}_2) \cdot \hat{B}(\alpha_1', \vec{\theta}_1).
\end{align*}

For our purposes, however, it is more practical to put the mild constraint that the sum of all displacements vanishes, \abbr{ie} $\alpha_{T+1}'=\sum_{s=1}^{T+1}\alpha_s\stackrel{!}{=}0$. This condition fixes only one degree of freedom of the gate sequence, and hence does not take away much of its expressive power.

\subsection{Length estimate}
\label{sec:appendix:gate_seq:length_estimate}

Here, we provide more details on the statement in the introduction that the optimum sequence length should scale linear with the number of involved Fock states.

This statement refers primarily to the situation where an entire Fock subspace is transformed, \abbr{ie} if the input space coincides with the output space, and this subspace is the span of Fock states (and not just embedded in the span of all Fock states with non-vanishing overlap), like for the examples in \cref{fig:results_unictrl}. If this is not the case, like for the examples in \cref{fig:results_applications}, a length estimate becomes more difficult. For instance, also the dimension of the logical subspace will play a rule then. In this case, the estimate for the transformation of the entire Fock subspace needs to be understood as an upper bound.

Focusing on the simple situation, our parameter-counting argument, here in more detail, works as follows: Searching for a gate sequence to realize a target operation can be interpreted as solving a high-dimensional (non-linear) system of equations. We can find a solution if the number of constraints does not exceed the number of free variables. There are as many constraints as angles to parameterize an operation on $N$ Fock states, which is $\dim(\mathset{SU}(N))=N^2-1=\mathcal{O}(N^2)$. The free variables are the control parameters of the gate sequence. The control parameters for a SNAP gate are the phase shifts for all the Fock levels, so, in principle, there are infinitely many of them. In practice, however, only the $\mathcal{O}(N)$ Fock levels close to the input/output space will get populated significantly, so effectively each SNAP gate contributes only $\mathcal{O}(N)$ free variables. In addition, each displacement contributes one more free parameter (the absolute of $\alpha$, as the complex phase can be absorbed into the neighboring SNAP gates, see \cref{sec:appendix:gate_seq:ansatz}). So, in total, we obtain $\mathcal{O}(N)$ free variables per pair of SNAP gate and displacement. Hence, we need $\mathcal{O}(N^2)/\mathcal{O}(N)=\mathcal{O}(N)$ pairs of these gates to arrive at a sequence with sufficiently many control parameters.

This optimistic estimate, promising quadratic speedup over the previous technique from \cite{krastanov2015snap_theory}, was actually the starting point for this project. It motivated us to look deeper into optimizing this control problem of combining SNAP gates and displacements, which has eventually led to the technique described in this paper.

\section{Objective functions}

\subsection{Overview}
\label{sec:appendix:objectives:summary}

To quantify the mismatch between the target operation and the actual transformation of the cavity state (disregarding noise and gate imperfections), we consider the mean overlap
\begin{align*}
	\quant{fidelity}[\hat{U}] & :=
		\max_{\varphi\in\mathset{R}} \avg_{\substack{\text{$\vec{c}\in\mathset{C}^L$ \abbr{st}}\\\sum_l\abs{c_l}^2=1}} \Real{\Braket{y(\vec{c})}{\mconst{e}^{\mconst{i}\varphi}\hat{U}}{x(\vec{c})}} = \\
	& =
		\frac{1}{L}\abs[\big]{\hsprod{\hat{V}}{\hat{U}}}
\end{align*}
as defined in \cref{eq:def_mean_overlap}. For $\hat{U}$, insert $\hat{U}= \hat{D}(\alpha_{T+1})\cdot\hat{S}(\vec{\theta}_T)\cdot\hat{D}(\alpha_T)\cdot \hdots\cdot\hat{S}(\vec{\theta}_2)\cdot\hat{D}(\alpha_2)\cdot\hat{S}(\vec{\theta}_1)\cdot\hat{D}(\alpha_1)$ as in \cref{eq:def_gate_sequence}, or its reparameterization $\hat{U}=\hat{B}_T\cdot\hdots\cdot\hat{B}_1$ using the building blocks defined in \cref{eq:def_building_blocks}. The interpretation as the mean overlap gets obvious from the first line. The second line, using the Hilbert-Schmidt product $\hsprod{\hat{A}}{\hat{B}}=\tr[\hat{A}^\dagger\hat{B}]$, is more useful to perform calculations. Their equivalence is shown in \cref{sec:appendix:objectives:rewrite_mean_overlap}.

Another important property of the gate sequences is the mean photon number, which determines the rate of photon losses. For this purpose, we introduce $\bar{n}_t$ as the mean photon number during $\hat{S}(\vec{\theta}_t)$, \abbr{ie} after executing $\hat{D}(\alpha_{t-1})\cdot\hat{S}(\vec{\theta}_{t-2})\cdot\hdots\cdot\hat{D}(\alpha_2)\cdot\hat{S}(\vec{\theta}_1)\cdot\hat{D}(\alpha_1)$, averaged over all input states $\ket{x(\vec{c})}$.

To obtain an expression for the photon number cost that is more symmetric than $\sum_{t=1}^{T}\bar{n}_t$, we follow the idea that mapping the input states $\ket{x(\vec{c})}$ to the output states $\ket{y(\vec{c})}$ with the gate sequence $\hat{D}(\alpha_{T+1})\cdot\hat{S}(\vec{\theta}_T)\cdot\hat{D}(\alpha_T)\cdot\hdots\cdot\hat{S}(\vec{\theta}_2)\cdot\hat{D}(\alpha_2)\cdot\hat{S}(\vec{\theta}_1)\cdot\hat{D}(\alpha_1)$ is equivalent to mapping the output states $\ket{y(\vec{c})}$ to the input states $\ket{x(\vec{c})}$ with the inverse sequence $\hat{D}^\dagger(\alpha_1)\cdot\hat{S}^\dagger(\vec{\theta}_1)\cdot\hat{D}^\dagger(\alpha_2)\cdot\hat{S}^\dagger(\vec{\theta}_2)\cdot\hdots\cdot\hat{D}^\dagger(\alpha_T)\cdot\hat{S}^\dagger(\vec{\theta}_T)\cdot\hat{D}^\dagger(\alpha_{T+1})$. Therefore, we minimize $\sum_{t=1}^{T}\frac{\bar{n}_t+\bar{n}_t'}{2}$ rather than $\sum_{t=1}^{T}\bar{n}_t$, where $\bar{n}_t'$ is the mean photon number after executing $\hat{D}^\dagger(\alpha_t)\cdot\hat{S}^\dagger(\vec{\theta}_{t+2})\cdot\hdots\cdot\hat{D}^\dagger(\alpha_T)\cdot\hat{S}^\dagger(\vec{\theta}_T)\cdot\hat{D}^\dagger(\alpha_{T+1})$, starting from the output states $\ket{y(\vec{c})}$.

The quantities $\bar{n}_t$ and $\bar{n}_t'$ can be computed as
\begin{align*}
	\bar{n}_t & :=
		\avg_{\substack{\text{$\vec{c}\in\mathset{C}^L$ \abbr{st}}\\\sum_l\abs{c_l}^2=1}} \Braket{x_t(\vec{c})}{\hat{n}}{x_t(\vec{c})} = \\
	& =
		\frac{1}{L}
		\hsprod{
			\hat{V}^\dagger\hat{V}
		}{
			\hat{B}_1^\dagger \cdot \hdots \cdot \hat{B}_{t-1}^\dagger
			\cdot
			\hat{D}(\alpha_t) \cdot \hat{n} \cdot \hat{D}^\dagger(\alpha_t)
			\cdot
			\hat{B}_{t-1} \cdot \hdots \cdot \hat{B}_1
		} \\
	\bar{n}_t' & :=
		\avg_{\substack{\text{$\vec{c}\in\mathset{C}^L$ \abbr{st}}\\\sum_l\abs{c_l}^2=1}} \Braket{y_t(\vec{c})}{\hat{n}}{y_t(\vec{c})} = \\
	& =
		\frac{1}{L}
		\hsprod{
			\hat{V}^\dagger\hat{V}
		}{
			\hat{B}_T \cdot \hdots \cdot \hat{B}_{t+1}
			\cdot
			\hat{D}(\alpha_t) \cdot \hat{n} \cdot \hat{D}^\dagger(\alpha_t)
			\cdot
			\hat{B}_{t+1}^\dagger \cdot \hdots \cdot \hat{B}_T^\dagger
		}
\end{align*}
with $\ket{x_t(\vec{c})}=\hat{D}(\alpha_t)\cdot\hat{S}(\vec{\theta}_{t-1})\cdot\hdots\cdot\hat{D}(\alpha_2)\cdot\hat{S}(\vec{\theta}_1)\cdot\hat{D}(\alpha_1)\ket{x(\vec{c})}=\hat{D}^\dagger(\alpha_t)\cdot\hat{B}_{t-1}\cdot\hdots\cdot\hat{B}_1\ket{x(\vec{c})}$ and $\ket{y_t(\vec{c})}=\hat{D}^\dagger(\alpha_t)\cdot\hat{B}_{t+1}^\dagger\cdot\hdots\cdot\hat{B}_T^\dagger\ket{y(\vec{c})}$.

\subsection{Calculation for the mean overlap}
\label{sec:appendix:objectives:rewrite_mean_overlap}

Here, we demonstrate how the definition for $\quant{fidelity}[\hat{U}]$ can be rewritten into a more practical form. We start by evaluating the maximization over $\varphi$:
\begin{align*}
	\quant{fidelity}[\hat{U}] & =
		\max_{\varphi\in\mathset{R}} \avg_{\substack{\text{$\vec{c}\in\mathset{C}^L$ \abbr{st}}\\\sum_l\abs{c_l}^2=1}} \Real[\big]{\Braket{y(\vec{c})}{\mconst{e}^{\mconst{i}\varphi}\hat{U}}{x(\vec{c})}} = \\
	& =
		\max_{\varphi\in\mathset{R}} \Real[\Big]{\mconst{e}^{\mconst{i}\varphi}\avg_{\substack{\text{$\vec{c}\in\mathset{C}^L$ \abbr{st}}\\\sum_l\abs{c_l}^2=1}} \Braket{y(\vec{c})}{\hat{U}}{x(\vec{c})}} = \\
	& =
		\abs[\Big]{\avg_{\substack{\text{$\vec{c}\in\mathset{C}^L$ \abbr{st}}\\\sum_l\abs{c_l}^2=1}} \Braket{y(\vec{c})}{\hat{U}}{x(\vec{c})}}
\end{align*}

By inserting $\ket{x(\vec{c})}=\sum_{l=1}^{L}c_l\ket{x(\unitvec{_l})}$ and $\ket{y(\vec{c})}=\sum_{l=1}^{L}c_l\ket{y(\unitvec{_l})}$, we can further simplify this expression:
\begin{align*}
	\quant{fidelity}[\hat{U}] & =
		\abs[\Big]{\avg_{\substack{\text{$\vec{c}\in\mathset{C}^L$ \abbr{st}}\\\sum_l\abs{c_l}^2=1}} \Braket{y(\vec{c})}{\hat{U}}{x(\vec{c})}} = \\
	& =
		\abs[\Big]{
			\sum_{l,l'=1}^{L} ~
			\overbrace{
				\avg_{\substack{\text{$\vec{c}\in\mathset{C}^L$ \abbr{st}}\\\sum_l\abs{c_l}^2=1}}[\conj{c_l}c_{l'}]
			}^{
				= \delta_{ll'}/L
			}
			\Braket{y(\unitvec{_l})}{\hat{U}}{x(\unitvec{_{l'}})}
		} = \\
	& =
		\frac{1}{L}
		\abs[\Big]{
			\sum_{l=1}^{L}
			\Braket{y(\unitvec{_l})}{\hat{U}}{x(\unitvec{_l})}
		} = \\
	& =
		\frac{1}{L}
		\abs[\bigg]{\tr\Big[
			\underbrace{
				\Big(\sum_{l=1}^{L} \ketbra{x(\unitvec{_l})}{y(\unitvec{_l})}\Big)
			}_{
				=\hat{V}^\dagger
			}
			\hat{U}
		\Big]} = \\
	& =
		\frac{1}{L} \abs[\big]{\hsprod{\hat{V}}{\hat{U}}}
\end{align*}

\section{Initialization}
\label{sec:appendix:initialization}

In \cref{sec:technique:initialization}, we describe a greedy strategy to initialize a gate sequence. Each step in this approach can be reduced to determining the optimum combination of $\alpha$ and $\vec{\theta}$ that maximizes mean overlap $\quant{fidelity}[\hat{B}(\alpha,\vec{\theta})]$ for a single building block $\hat{B}(\alpha,\vec{\theta})=\hat{D}^\dagger(\alpha)\hat{S}(\vec{\theta})\hat{D}(\alpha)$. Here, we derive the result as stated in \cref{sec:technique:initialization}.

First, we compute
\begin{equation*}
	\quant{fidelity}[\hat{B}(\alpha,\vec{\theta})] =
	\frac{1}{L} \abs{\hsprod{\hat{V}}{\hat{B}(\alpha,\vec{\theta})}} =
	\hdots =
	\frac{1}{L} \abs[\bigg]{
		\sum_{n=0}^{\infty}
		\mconst{e}^{\mconst{i}\theta^{(n)}}
		\conj{g_n}(\alpha)
	}
\end{equation*}
with $g_n(\alpha)=\Braket{n}{\hat{D}(\alpha)\hat{V}\hat{D}^\dagger(\alpha)}{n}$ as defined in \cref{sec:technique:initialization}. Now, we can optimize $\vec{\theta}$ for an arbitrary, but fixed value for $\alpha$:
\begin{align*}
	\argmax_{\theta^{(n)}} \quant{fidelity}[\hat{B}(\alpha,\vec{\theta})] & =
		\bar{\theta} + \arg(g_n(\alpha)) + 2\mconst{pi}\mathset{Z} \\
	\max_{\vec{\theta}} \quant{fidelity}[\hat{B}(\alpha,\vec{\theta})] & =
		\frac{1}{L} \sum_{n=0}^{\infty} \abs{g_n(\alpha)}.
\end{align*}
Hence, the optimum value for $\alpha$ is
\begin{equation*}
	\alpha_\mathrm{opt} = \argmax_{\alpha} \bigg[\sum_{n=0}^{\infty} \abs{g_n(\alpha)}\bigg],
\end{equation*}
and the optimum value for $\vec{\theta}$ is $\vec{\theta}_\mathrm{opt}=\bar{\theta}+\arg(g_n(\alpha_\mathrm{opt}))+2\mconst{pi}\mathset{Z}$. By choosing $\bar{\theta}=0$, we arrive at \cref{eq:initialization} in the main text.

This recipe can be used to extend a sequence of building blocks $\hat{B}(\alpha,\vec{\theta})$ to any desired length in a meaningful way, as explained in \cref{sec:technique:initialization}.

\section{Finetuning}

During finetuning, we take the gradient of the cost function in \cref{eq:finetuning_cost_function} \abbr{wrt} the control parameters $\alpha_t$ and $\vec{\theta}_t$ of the building blocks $\hat{B}_t=\hat{B}(\alpha_t,\vec{\theta}_t)$. Here, we provide explicit expressions for the gradients. We also demonstrate how these gradients can be computed efficiently using a recursive scheme which is analogous to backpropagation in machine learning.

\subsection{Gradient for the overlap cost}
\label{sec:appendix:finetuning:overlap}

The gradient for the overlap cost is given by
\begin{align*}
	\frac{\partial}{\partial\alpha_t} \ln\bigg(1-\quant{fidelity}[\hat{B}_T\cdot\hdots\cdot\hat{B}_1]\bigg) & =
		\Real[\Bigg]{
			\hsprod[\bigg]{
				\hat{G}_t
			}{
				\frac{\partial\hat{B}(\alpha_t,\vec{\theta}_t)}{\partial\alpha_t}
			}
		} \\
	\frac{\partial}{\partial\theta_t^{(n)}} \ln\bigg(1-\quant{fidelity}[\hat{B}_T\cdot\hdots\cdot\hat{B}_1]\bigg) & =
		\Real[\Bigg]{
			\hsprod[\bigg]{
				\hat{G}_t
			}{
				\frac{\partial\hat{B}(\alpha_t,\vec{\theta}_t)}{\partial\theta_t^{(n)}}
			}
		}
\end{align*}
with
\begin{equation*}
	\hat{G}_t :=
	-
	\frac{
		\hsprod{\hat{V}}{\hat{B}_T\cdot\hdots\cdot\hat{B}_1}
	}{
		\abs{\hsprod{\hat{V}}{\hat{B}_T\cdot\hdots\cdot\hat{B}_1}}
	}
	\cdot
	\frac{
		\hat{B}_{t+1}^\dagger \cdot \hdots \cdot \hat{B}_T^\dagger
		\cdot
		\hat{V}
		\cdot
		\hat{B}_1^\dagger \cdot \hdots \cdot \hat{B}_{t-1}^\dagger
	}{
		L-\abs{\hsprod{\hat{V}}{\hat{B}_T\cdot\hdots\cdot\hat{B}_1}}
	}.
\end{equation*}

$\hat{G}_t$ can be efficiently computed using the recursive scheme
\begin{align*}
	\hat{G}_1 & =
		-
		\frac{
			\hsprod{\hat{V}}{\hat{B}_T\cdot\hdots\cdot\hat{B}_1}
		}{
			\abs{\hsprod{\hat{V}}{\hat{B}_T\cdot\hdots\cdot\hat{B}_1}}
		}
		\cdot
		\frac{
			\hat{B}_2^\dagger \cdot \hdots \cdot \hat{B}_T^\dagger \cdot \hat{V}
		}{
			L-\abs{\hsprod{\hat{V}}{\hat{B}_T\cdot\hdots\cdot\hat{B}_1}}
		} &
	\hat{G}_{t+1} & =
		\hat{B}_{t+1} \hat{G}_t \hat{B}_t^\dagger.
\end{align*}

\subsection{Gradient for the photon-number cost}

The gradient for the photon-number cost can be computed from
\begin{align*}
	\frac{\partial}{\partial\alpha_t}\bigg[\sum_{s=1}^{T}\bar{n}_s\bigg] & =
		\Real[\Bigg]{
			\hsprod[\bigg]{
				2\hat{X}_t\hat{B}_t\hat{\rho}_t^\mathrm{(x)}
			}{
				\frac{\partial\hat{B}(\alpha_t,\vec{\theta}_t)}{\partial\alpha_t}
			} +
			\hsprod[\bigg]{
				2\hat{\rho}_t^\mathrm{(x)}
			}{
				\hat{D}^\dagger(\alpha_t) \hat{n} \frac{\partial\hat{D}(\alpha_t)}{\partial\alpha_t}
			}
		} \\
	\frac{\partial}{\partial\theta_t^{(n)}}\bigg[\sum_{s=1}^{T}\bar{n}_s\bigg] & =
		\Real[\Bigg]{
			\hsprod[\bigg]{
				2\hat{X}_t\hat{B}_t\hat{\rho}_t^\mathrm{(x)}
			}{
				\frac{\partial\hat{B}(\alpha_t,\vec{\theta}_t)}{\partial\theta_t^{(n)}}
			}
		} \\
	\frac{\partial}{\partial\alpha_t}\bigg[\sum_{s=1}^{T}\bar{n}_s'\bigg] & =
		\Real[\Bigg]{
			\hsprod[\bigg]{
				2\hat{\rho}_t^\mathrm{(y)}\hat{B}_t\hat{Y}_t
			}{
				\frac{\partial\hat{B}(\alpha_t,\vec{\theta}_t)}{\partial\alpha_t}
			} +
			\hsprod[\bigg]{
				2\hat{\rho}_t^\mathrm{(y)}
			}{
				\hat{D}^\dagger(\alpha_t) \hat{n} \frac{\partial\hat{D}(\alpha_t)}{\partial\alpha_t}
			}
		} \\
	\frac{\partial}{\partial\theta_t^{(n)}}\bigg[\sum_{s=1}^{T}\bar{n}_s'\bigg] & =
		\Real[\Bigg]{
			\hsprod[\bigg]{
				2\hat{\rho}_t^\mathrm{(y)}\hat{B}_t\hat{Y}_t
			}{
				\frac{\partial\hat{B}(\alpha_t,\vec{\theta}_t)}{\partial\theta_t^{(n)}}
			}
		}.
\end{align*}
with
\begin{align*}
	\hat{\rho}_t^\mathrm{(x)} & =
		\frac{1}{L} \cdot
		\hat{B}_{t-1} \cdot \hdots \cdot \hat{B}_1
		\cdot
		\hat{V}^\dagger \cdot \hat{V}
		\cdot
		\hat{B}_1^\dagger \cdot \hdots \cdot \hat{B}_{t-1}^\dagger \\
	\hat{\rho}_t^\mathrm{(y)} & =
		\frac{1}{L} \cdot
		\hat{B}_{t+1}^\dagger \cdot \hdots \cdot \hat{B}_T^\dagger
		\cdot
		\hat{V} \cdot \hat{V}^\dagger
		\cdot
		\hat{B}_T \cdot \hdots \cdot \hat{B}_{t+1} \\
	\hat{X}_t & =
		\sum_{s=t+1}^T
		\hat{B}_{t+1}^\dagger \cdot \hdots \cdot \hat{B}_{s-1}^\dagger
		\cdot
		\hat{D}^\dagger(\alpha_s) \cdot \hat{n} \cdot \hat{D}(\alpha_s)
		\cdot
		\hat{B}_{s-1} \cdot \hdots \cdot \hat{B}_{t+1} \\
	\hat{Y}_t & =
		\sum_{s=1}^{t-1}
		\hat{B}_{t-1} \cdot \hdots \cdot \hat{B}_{s+1}
		\cdot
		\hat{D}^\dagger(\alpha_s) \cdot \hat{n} \cdot \hat{D}(\alpha_s)
		\cdot
		\hat{B}_{s+1}^\dagger \cdot \hdots \cdot \hat{B}_{t-1}^\dagger.
\end{align*}

Again, there are recursive schemes to compute the quantities $\hat{\rho}_t^\mathrm{(x)}$, $\hat{\rho}_t^\mathrm{(y)}$, $\hat{X}_t$ and $\hat{Y}_t$ efficiently:
\begin{align*}
	\hat{\rho}_1^\mathrm{(x)} & = \frac{1}{L} \hat{V}^\dagger \hat{V} &
		\hat{\rho}_{t+1}^\mathrm{(x)} & = \hat{B}_t \hat{\rho}_t^\mathrm{(x)} \hat{B}_t^\dagger \\
	\hat{\rho}_T^\mathrm{(y)} & = \frac{1}{L} \hat{V} \hat{V}^\dagger &
		\hat{\rho}_{t-1}^\mathrm{(y)} & = \hat{B}_t^\dagger \hat{\rho}_t^\mathrm{(y)} \hat{B}_t \\
	\hat{X}_T & = 0 &
		\hat{X}_{t-1} & = \hat{B}_t^\dagger \hat{X}_t \hat{B}_t + \hat{D}^\dagger(\alpha_t) \hat{n} \hat{D}(\alpha_t) \\
	\hat{Y}_1 & = 0 &
		\hat{Y}_{t+1} & = \hat{B}_t \hat{Y}_t \hat{B}_t^\dagger + \hat{D}^\dagger(\alpha_t) \hat{n} \hat{D}(\alpha_t)
\end{align*}

\section{Benchmarks}

\subsection{Transformations of the 10 lowest Fock states}
\label{sec:appendix:examples:unictrl}

Here, we provide explicit formulas for the operators $\hat{V}$ representing the target operation (\abbr{cmp} \cref{eq:def_gate_sequence}), for all examples in \cref{fig:results_unictrl}a.

\noindent \textbf{Inversion:} $\hat{V}=\sum_{n=0}^{9}\ketbra{9-n}{n}$

\noindent \textbf{Block inversion:} $\hat{V}=\sum_{n=0}^{9}\ketbra{\mathrm{mod}(n+5,10)}{n}$

\noindent \textbf{Random permutations:} $\hat{V} = \sum_{n=0}^{9} \ketbra{p(n)}{n}$ where $p$ are random permutations of the numbers $0,\hdots,9$.

\noindent \textbf{Random unitaries:} $\hat{V} = \sum_{m,n=0}^{9} v_{mn} \ketbra{m}{n}$ where $v$ is a unitary $10\times10$ matrix generated by diagonalizing random Hermitian matrices.

\subsection{Applied problems}
\label{sec:appendix:examples:applied}

Here, we provide explicit formulas for the operators $\hat{V}$ representing the target operation (\abbr{cmp} \cref{eq:def_gate_sequence}), for all examples in \cref{fig:results_applications}.

\noindent \textbf{State preparation} (\cref{fig:results_applications}a):
\begin{equation*}
	\hat{V} = (\alpha\ket{b_0}+\beta\ket{b_1}) \bra{0}
\end{equation*}
for the stated values of the coefficients $\alpha$ and $\beta$. The numbers for ``odd superposition'' are $\alpha=\sqrt{(1+\sin0.72104)/2}$ and $\beta=\sqrt{(1-\sin0.72104)/2}\cdot\mconst{e}^{-1.27275\mconst{i}}$.

\noindent \textbf{Error correction} (\cref{fig:results_applications}b):
\begin{equation*}
	\hat{V}_s = \ketbra[\big]{\psi_{1,0}^{(\unitmat)}(0)}{\psi_{1,0}^{(s)}(t)} + \ketbra[\big]{\psi_{0,1}^{(\unitmat)}(0)}{\psi_{0,1}^{(s)}(t)}
\end{equation*}
with $\Gamma t=0.02$ and
\begin{align*}
	\ket{\psi_{\alpha\beta}^{(\unitmat)}(t)} & =
		\frac{
			\alpha\cdot\big(\ket{0}+\sqrt{3}\mconst{e}^{-6\Gamma t}\ket{6}\big)
			+
			\beta\cdot\big(\sqrt{3}\mconst{e}^{-3\Gamma t}\ket{3}+\mconst{e}^{-9\Gamma t}\ket{9}\big)
		}{\sqrt{
			\abs{\alpha}^2\cdot(1+3\mconst{e}^{-12\Gamma t})
			+
			\abs{\beta}^2\cdot(3\mconst{e}^{-6\Gamma t}+\mconst{e}^{-18\Gamma t})
		}} \\
	\ket{\psi_{\alpha\beta}^{(\hat{a})}(t)} & =
		\frac{
			\alpha\cdot\big(\sqrt{2}\mconst{e}^{-3\Gamma t}\ket{5}\big)
			+
			\beta\cdot\big(\ket{2}+\mconst{e}^{-6\Gamma t}\ket{8}\big)
		}{\sqrt{
			\abs{\alpha}^2\cdot(2\mconst{e}^{-6\Gamma t})
			+
			\abs{\beta}^2\cdot(1+\mconst{e}^{-12\Gamma t})
		}} \\
	\ket{\psi_{\alpha\beta}^{(\hat{a}^2)}(t)} & =
		\frac{
			\alpha\cdot\big(\sqrt{5}\mconst{e}^{-3\Gamma t}\ket{4}\big)
			+
			\beta\cdot\big(\ket{1}+2\mconst{e}^{-6\Gamma t}\ket{7}\big)
		}{\sqrt{
			\abs{\alpha}^2\cdot(5\mconst{e}^{-6\Gamma t})
			+
			\abs{\beta}^2\cdot(1+4\mconst{e}^{-12\Gamma t})
		}}
\end{align*}
These states can be computed as follows: To solve the Lindblad master equation for a cavity under photon loss,
\begin{equation*}
	\frac{\total{\hat{\rho}_{\alpha\beta}(t)}}{\total{t}} = \Gamma \cdot \big(2\hat{a}\hat{\rho}_{\alpha\beta}(t)\hat{a}^\dagger-\anticomm{\hat{a}^\dagger\hat{a}}{\hat{\rho}_{\alpha\beta}(t)}\big),
\end{equation*}
with initial state $\hat{\rho}_{\alpha\beta}(0)=(\alpha\ket{b_0}+\beta\ket{b_1})(\alpha\ket{b_0}+\beta\ket{b_1})^\dagger$, we decompose $\hat{\rho}_{\alpha\beta}(t)$ as $\hat{\rho}_{\alpha\beta}(t)=\sum_{j=0}^{9}\hat{\rho}_{\alpha\beta}^{(\hat{a}^j)}(t)$ with
\begin{align*}
	&&
	\frac{1}{\Gamma} \frac{\total{\hat{\rho}_{\alpha\beta}^{(\unitmat)}(t)}}{\total{t}} & =
		- \anticomm{\hat{n}}{\hat{\rho}_{\alpha\beta}^{(\unitmat)}(t)} &
	\hat{\rho}_{\alpha\beta}^{(\unitmat)}(0) & =
		\unitmat \\
	& j\ge1: &
	\frac{1}{\Gamma} \frac{\total{\hat{\rho}_{\alpha\beta}^{(\hat{a}^j)}(t)}}{\total{t}} & =
		- \anticomm{\hat{n}}{\hat{\rho}_{\alpha\beta}^{(\hat{a}^j)}(t)} + 2 \hat{a} \hat{\rho}_{\alpha\beta}^{(\hat{a}^{j-1})}(t) \hat{a}^\dagger &
	\hat{\rho}_{\alpha\beta}^{(\hat{a}^j)}(0) & =
		0.
\end{align*}
The solutions $\hat{\rho}_{\alpha\beta}^{(\hat{a}^j)}(t)$ all have rank 1 (unless $t=0$). $\ket{\psi_{\alpha\beta}^{(\unitmat)}(t)}$, $\ket{\psi_{\alpha\beta}^{(\hat{a})}(t)}$ and $\ket{\psi_{\alpha\beta}^{(\hat{a}^2)}(t)}$ are chosen such that they satisfy $\Ketbra{\psi_{\alpha\beta}^{(\hat{a}^j)}(t)}=\hat{\rho}_{\alpha\beta}^{(\hat{a}^j)}(t)/\tr[\hat{\rho}_{\alpha\beta}^{(\hat{a}^j)}(t)]$. \\
Note that $\ket{\psi_{1,0}^{(\unitmat)}(0)}=\ket{b_0}$ and $\ket{\psi_{0,1}^{(\unitmat)}(0)}=\ket{b_1}$.

\noindent \textbf{Logical operations on binomial code} (\cref{fig:results_applications}c):
\begin{equation*}
	\hat{V} = \sum_{j,k=0}^{1} v_{jk} \ketbra{b_j}{b_k}
\end{equation*}
with
\begin{center}
	\begin{tabular}{ccccc}
		$v=\frac{1}{\sqrt{2}}\begin{pmatrix}1&1\\1&-1\end{pmatrix}$ &
		$v=\begin{pmatrix}0&1\\1&0\end{pmatrix}$ &
		$v=\begin{pmatrix}0&-\mconst{i}\\\mconst{i}&0\end{pmatrix}$ &
		$v=\frac{1}{2}\begin{pmatrix}1-\mconst{i}&1+\mconst{i}\\1+\mconst{i}&1-\mconst{i}\end{pmatrix}$ \\
		(Hadamard) & (Pauli-X) & (Pauli-Y) & (Sqrt-Pauli-X)
	\end{tabular}
\end{center}
$v$ for ``odd operation'' is a $2\times2$ matrix that was generated by diagonalizing a random Hermitian matrix.

\noindent \textbf{Logical operations on trivial code} (\cref{fig:results_applications}d):
\begin{equation*}
	\hat{V} = \sum_{m,n=0}^{1} v_{mn} \ketbra{m}{n}
\end{equation*}
with the same $v$ matrices as for \cref{fig:results_applications}c.

\subsection{Comparison with previous technique}
\label{sec:appendix:examples:comparison}

The following table compares the sequence length between our technique and the technique in \cite{krastanov2015snap_theory}, for the target operations in \cref{fig:results_applications}. For \cite{krastanov2015snap_theory}, we give the minimum sequence lengths, which is the squared number of involved Fock states. For our technique, we state the smallest value for $T$ such that $\quant{fidelity}$ exceeds a value of $0.999$.

\begin{center}
	\begin{tabular}{|l|l|l|r|r|}
		\hline
		\multicolumn{2}{|c|}{target operation} & \begin{tabular}{@{}l@{}}involved\\Fock levels\end{tabular} & $T_\mathrm{min}$ for \cite{krastanov2015snap_theory} & $T_{\quant{fidelity}\ge0.999}$ here \\ \hline
		\hline
		\multirow{3}{*}{\begin{tabular}{@{}l@{}}state preparation\\(\cref{fig:results_applications}a)\end{tabular}} &
			$\ket{b_0}$ & $0,6$ & $4$ & $4$ \\
		&
			$\ket{b_1}$ & $0,3,9$ & $9$ & $3$ \\
		&
			superpositions & $0,3,6,9$ & $16$ & $3\hdots4$ \\ \hline
		\multirow{3}{*}{\begin{tabular}{@{}l@{}}error correction\\(\cref{fig:results_applications}b)\end{tabular}} &
			syndrome $\unitmat$ & $0,3,6,9$ & $16$ & \multirow{2}{*}{$4$} \\
		&
			syndrome $\hat{a}$ & $0,2,3,5,6,8,9$ & $49$ & \\
		&
			syndrome $\hat{a}^2$ & $0,1,3,4,6,7,9$ & $49$ & \\ \hline
		\begin{tabular}{@{}l@{}}logical operations on\\binomial code (\cref{fig:results_applications}c)\end{tabular} &
			& $0,3,9$ & $16$ & $3\hdots4$ \\ \hline
		\begin{tabular}{@{}l@{}}logical operations on\\trivial code (\cref{fig:results_applications}d)\end{tabular} &
			& $0,1$ & $4$ & $3$ \\ \hline
	\end{tabular}
\end{center}

\section{Simulations}

\subsection{Implementation details and hyperparameters}
\label{sec:appendix:simulations:hyperparams}


\noindent \textbf{Both during initialization and finetuning:}
\begin{itemize}
	\item The Hilbert space of a cavity is infinite-dimensional, so in order to be able to simulate it, we have to truncate it at some point. In all our simulations, we neglect those Fock states $\ket{n}$ with $n\ge100$. We choose this limit deliberately high to rule out any non-negligible leakage of the relevant input states out of this space, guaranteeing for correctness of our calculations. With $100$ simulated Fock states, leakage to higher states is much smaller than the numerical floating-point error, so our choice is very generous. This goes at the expense of the simulation runtime (which scales quadratically with the truncation limit), but this we can afford.
\end{itemize}

\noindent \textbf{During initialization only:}
\begin{itemize}
	\item To find the optimal values for the control parameters $\alpha$ and $\vec{\theta}$ of an inserted building block according to \cref{eq:initialization}, we have to take into account in principle every real number for $\alpha$. In our implementation, we restrict our search to a fixed set of candidates for $\alpha$. Explicitly, we choose this set as $\{-2,-1.8,\hdots,1.8,2\}$. Intentionally, this set excludes values with $\abs{\alpha}$ above a certain threshold (here $2$), as large displacements lead to high values for the average mean photon number $\bar{n}_t$ (which we want to avoid).
	\item All SNAP phase angles for the Fock levels $n\ge15$ are set to $0$ during initialization. Even though this is not absolutely necessary, it helps to launch finetuning with a lower value for the average mean photon number $\bar{n}_t$. However, also during initialization, we simulate the cavity up to Fock state $n=100$ (see above) since also the Fock states $\ge15$ are still populated. Moreover, during finetuning, we allow non-vanishing SNAP phase angles also for the Fock states between $15$ and $99$ (even though in practice these phase angles do not change significantly during finetuning).
\end{itemize}

\noindent \textbf{During finetuning only:}
\begin{itemize}
	\item We run our finetuning procedure for $\SI{100000}{}$ iterations. The results in \cref{fig:results_unictrl} and \cref{fig:results_applications} always refer to the parameter configuration after the last iteration step. Due to fluctuations in the cost function, its value might have been slightly better at previous steps, but the difference is not large enough such that a post-selection over these intermediate configurations would make sense.
	\item We use the Adam optimizer \cite{kingma2014adam} to compute the updates for $\alpha$ and $\vec{\theta}$ from the corresponding gradients of the cost function. We set the Adam hyperparameters to $\eta=\SI{e-4}{}$, $\beta_1=0.9$, $\beta_2=0.999$ and $\epsilon=\SI{e-8}{}$.
	\item Gradient clipping is implemented by truncating the values for $\abs{\partial C/\partial\alpha_t}$ at $100.0$ and for $\abs{\partial C/\partial\theta_t^{(n)}}$ at $50.0$ before passing them to the Adam optimizer.
	\item The only hyperparameter which is changed between the different examples is $\lambda$, the coefficient for the photon number cost (see \cref{sec:technique:finetuning}). Explicitly, we use the following values: 
		
		\begin{center}
			\begin{tikzpicture}
				\node[align=left] (fig_3) at (0,0) {
					\cref{fig:results_applications}: \\
					\begin{tabular}{|l||l|l|}
						\hline
						application & $\lambda$ & conditions \\ \hline \hline
						state preparation (\cref{fig:results_applications}a) & $\lambda=0.6$ & * \\ \hline
						\multirow{2}{*}{error correction (\cref{fig:results_applications}b)} & $\lambda=0.6$ & syndrome $\hat{a}$ and $T\le4$ \\
						& $\lambda=0.4$ & otherwise \\ \hline
						\multirow{2}{*}{logical operation on binomial code (\cref{fig:results_applications}c)} & $\lambda=1.0$ & operation Pauli-X and $T\le3$ \\
						& $\lambda=0.32$ & otherwise \\ \hline
						logical operation on trivial code (\cref{fig:results_applications}d) & $\lambda=2.4$ & * \\ \hline
					\end{tabular}
				};
				\node[align=left,anchor=south west,shift={(0,0.5)}] (fig_2a) at (fig_3.north west) {
					\cref{fig:results_unictrl}a: \\
					\begin{tabular}{|l||c|c|}
						\hline
						target operation & $T\le8$ & $T\ge10$ \\ \hline \hline
						inversion & $\lambda=0.16$ & $\lambda=0.4$ \\ \hline
						block inversion & \multicolumn{2}{c|}{$\lambda=0.8$} \\ \hline
						random permutations & \multicolumn{2}{c|}{\multirow{2}{*}{$\lambda=1.6$}} \\ \cline{0-0}
						random unitaries & \multicolumn{2}{c|}{} \\ \hline
					\end{tabular}
				};
				\node[align=left,anchor=north east] at (fig_2a.north-|fig_3.east) {
					\cref{fig:results_unictrl}b: \\
					\begin{tabular}{|l|c|}
						\hline
						$N$ & $\lambda$ \\ \hline \hline
						$2,3$ & $\lambda=2.4$ \\
						$4,5$ & $\lambda=1.8$ \\
						$\ge6$ & $\lambda=1.6$ \\ \hline
					\end{tabular}
				};
			\end{tikzpicture}
		\end{center}
	\item  For the ``no GC, higher LR'' curve in \cref{fig:training}c, we  did not apply gradient clipping, and in addition changed two Adam hyperparameters: $\eta=\SI{2.5e-4}{}$ and $\beta_2=0.99$. For this parameter configuration, post-selection would be required to reliably find well-performing sequences. Therefore, we find our default hyperparameters preferable which avoid this additional step of complexity.
\end{itemize}

\subsection{Computational costs}

Because we need to construct a large number of gate sequences (for different target operations and sequence lengths), we can obtain the strongest acceleration from embarrassing parallelization, and therefore run each instance single-threaded on one CPU core. For each job, initialization is completed in less than $\SI{1}{\minute}$, and finetuning takes between $\SI{2}{\hour}$ and $\SI{40}{\hour}$ (for $\SI{100000}{}$ iterations). If needed, this runtime could potentially be reduced significantly with a more efficient implementation.

The RAM requirements are very modest. For the example with the highest memory usage, it takes $\SI{32}{\kibi\byte}$ to store the control parameters for one sequence, and between $\SI{10}{\mega\byte}$ and $\SI{20}{\mega\byte}$ to compute the gradient during finetuning.

The computations were distributed over 8 nodes \`{a} 32 cores.
\newpage


\begin{appendices}

\setcounter{figure}{0}
\renewcommand\thefigure{S\arabic{figure}}

\vspace*{5\baselineskip}
\begin{center}
	\LARGE{\textbf{Supplementary Material}}
\end{center}

\newpage

\begin{figure}
	\includegraphics[scale=0.6]{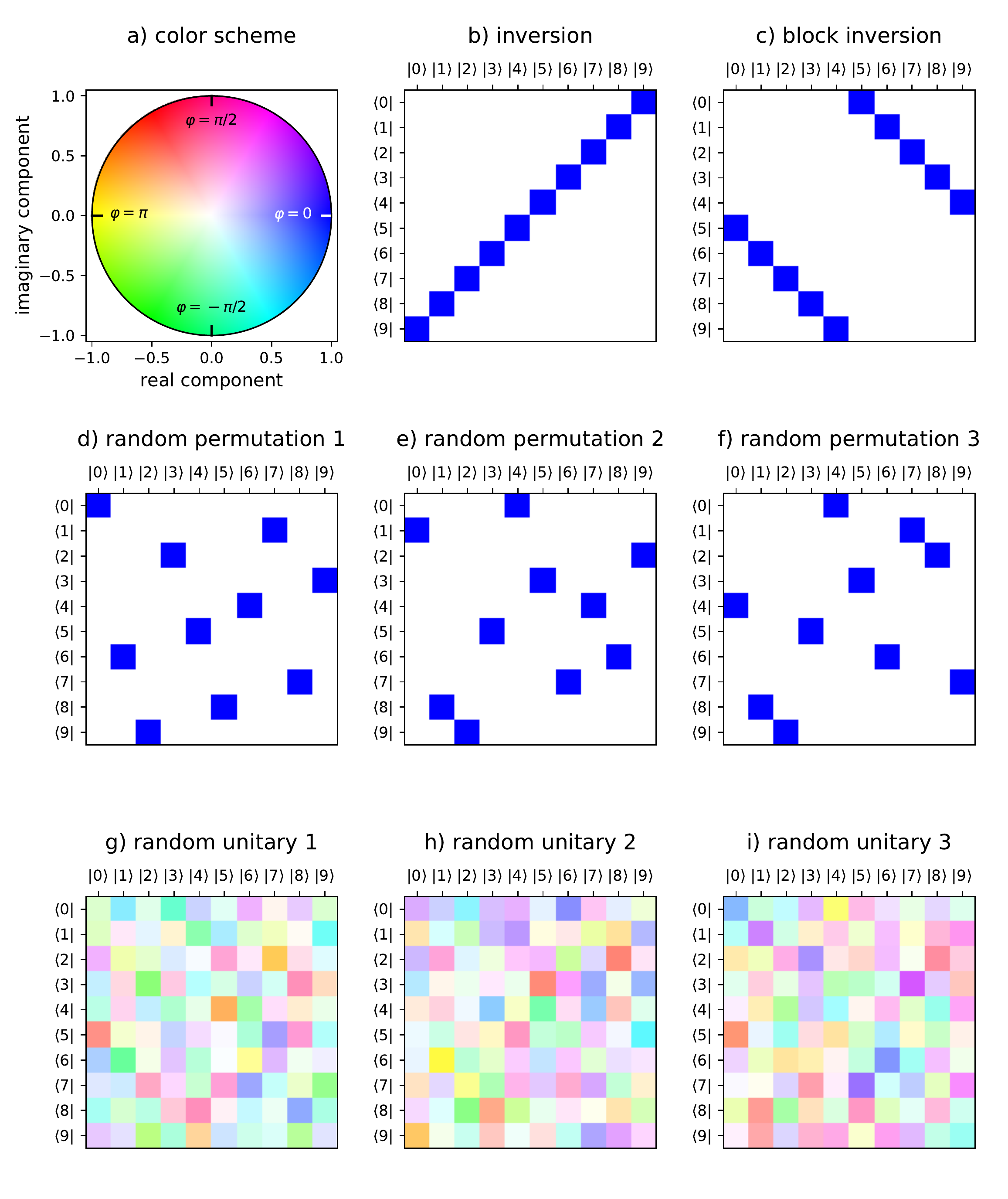}
	\caption{Supplementary information to \cref{fig:results_unictrl}a.
		a) Color scheme to represent complex numbers in the following subfigures.
		b-i) Graphical representation of the target operations used for \cref{fig:results_unictrl}a.
	}
\end{figure}

\begin{figure}
	\includegraphics[scale=0.6]{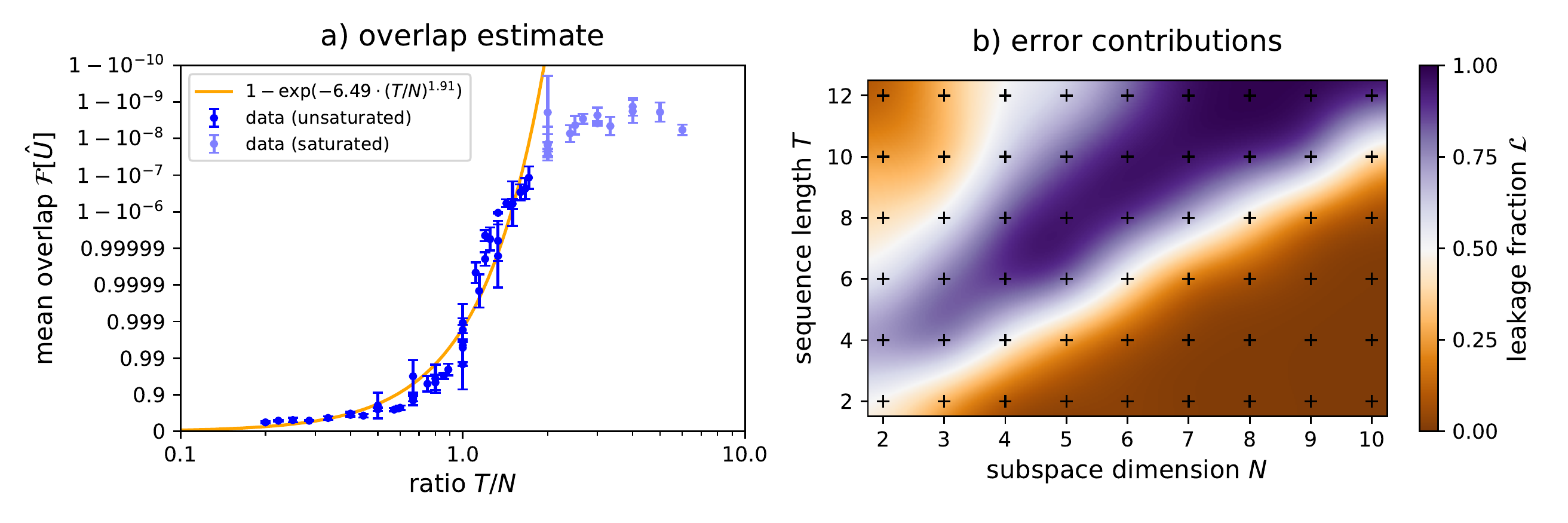}
	\caption{Supplementary information to \cref{fig:results_unictrl}b.
		a) Estimate for the mean overlap $\quant{fidelity}[\hat{U}]$ obtained for target operations on the $N$ lowest Fock states. Our result from \cref{fig:results_unictrl}b, that $T$, the number of building blocks in a sequence, should scale linear in $N$ to achieve the same mean overlap $\quant{fidelity}[\hat{U}]$, means that $\quant{fidelity}[\hat{U}]$ is primarily a function of the ratio $T/N$. Therefore, we plot here the averaged values for $\quant{fidelity}[\hat{U}]$ from \cref{fig:results_unictrl}b against $T/N$. We find that the data points indeed lie on a curve, up to small deviations. To get more insight into the dependency between $\quant{fidelity}[\hat{U}]$ and $T/N$, we make a fit to these data points; the orange curve shows the result. For this fit, we have left out the data points in the saturation region, where $\quant{fidelity}[\hat{U}]$ visibly follows a different behavior than for smaller values of $T/N$.
		b) Error contributions. The total error can be divided into two parts: leakage of the actual output states out of the desired output space, and a mismatch within this desired output space. To learn how strong these contributions are, we plot the value of $\mathcal{L}=(1-\quant{non_leakage}[\hat{U}])/(1-\quant{fidelity}[\hat{U}])$ into the same coordinate system as in \cref{fig:results_unictrl}b. Here, $\quant{non_leakage}[\hat{U}]=\frac{1}{L}\norm{\hat{V}\hat{U}^\dagger\hat{V}}_1$ quantifies the probability that the actual output states have not leaked out of the desired output space, averaged over all relevant input states. Hence, $\mathcal{L}$ gives the fraction to which leakage contributes to the total error. As the plot shows, leakage has a minor effect both for small and large values of $T/N$, whereas it is the dominating factor for intermediate values of $T/N$. This behavior could be a consequence of our particular technique to construct the gate sequences.
	}
	\label{fig:suppl_scaling_behavior}
\end{figure}

\begin{figure}
	\includegraphics[scale=0.6]{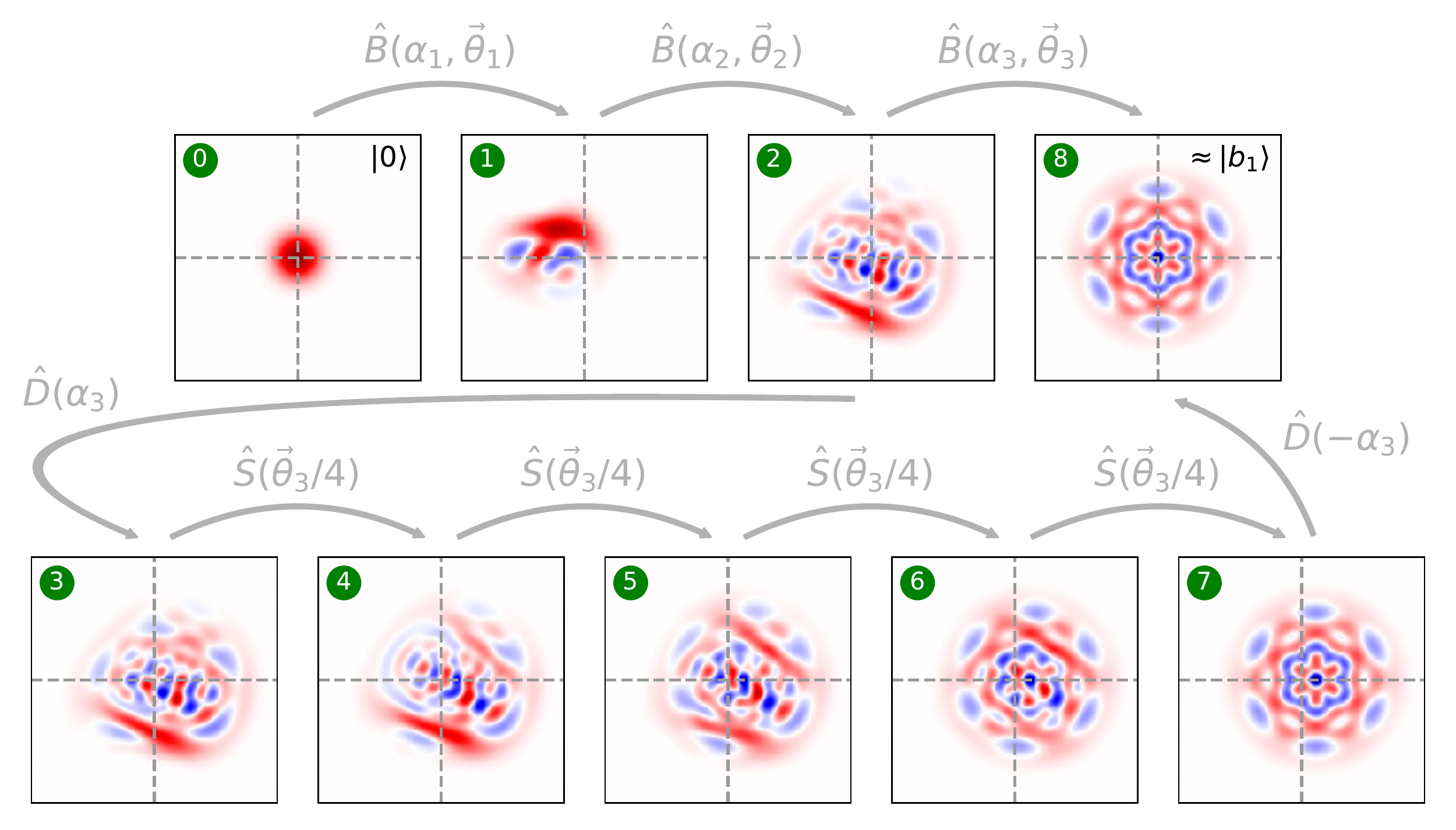}
	\caption{Supplementary information to \cref{fig:results_applications}e. The first row is equal to \cref{fig:results_applications}e, the second row shows additional Wigner density plots for the substeps of the final building block. The displacements ($2\to3$, $7\to8$) just shift the Wigner density along $x$, whereas the SNAP gate ($3\to\hdots\to7$) transforms the Wigner density in a complex manner.}
\end{figure}

\end{appendices}

\end{document}